\documentclass[amsmath,amsfonts,twocolumn,superscriptaddress,prb]{revtex4-2}

\usepackage[colorlinks=true,allcolors=blue,hypertexnames=false]{hyperref}

\usepackage{graphicx}
\usepackage{epsfig}
\usepackage{bbm}
\usepackage{bm}
\usepackage{dcolumn}
\usepackage{amsmath}
\usepackage{amssymb}
\usepackage{color}
\usepackage[varg]{txfonts}

\usepackage{graphicx}
\usepackage{epsfig}
\usepackage{bbm}
\usepackage{bm}
\usepackage{dcolumn}
\usepackage{amsmath}
\usepackage{amssymb}
\usepackage{color}
\usepackage[varg]{txfonts}
\usepackage{ulem}

\newcommand{\beg}{\begin{equation}}
\newcommand{\en}{\end{equation}}

\newcommand{\bq}{\mathbf q}
\newcommand{\bk}{\mathbf k}
\newcommand{\br}{\mathbf r}

\newcommand{\bB}{\mathbf B}
\newcommand{\bn}{\mathbf n}

\newcommand{\bj}{\mathbf j}

\newcommand \bel  {\begin{align}}
\newcommand \enl  {\end{align}}

\newcommand{\veps}{\varepsilon}
\newcommand{\eps}{\epsilon}

\newcommand{\dg}{^\dagger}

\newcommand{\vk}{\color{red}}

\def\XXint#1#2#3{{\setbox0=\hbox{$#1{#2#3}{\int}$}
     \vcenter{\hbox{$#2#3$}}\kern-.5\wd0}}

\begin{document}

\title{Nonlinear Hall effect in topological Dirac semimetals in parallel magnetic field}

\author{Maxim Dzero}
\affiliation{Department of Physics, Kent State University, Kent, Ohio 44242, USA}

\author{Maxim Khodas}
\affiliation{The Racah Institute of Physics, The Hebrew University of Jerusalem, Jerusalem 91904, Israel}

\author{Alex Levchenko}
\affiliation{Department of Physics, University of Wisconsin-Madison, Madison, Wisconsin 53706, USA}

\author{Vladyslav Kozii}
\affiliation{Department of Physics, Carnegie Mellon University, Pittsburgh, Pennsylvania 15213, USA}

\date{\today}

\begin{abstract}
We compute the second-harmonic response of two-dimensional topological Dirac semimetals subjected to an in-plane magnetic field.
The quantum kinetic equation for the Wigner distribution function is derived and then solved to compute the second-order electric-field contributions to the current density. Both the Berry curvature dipole and the field-induced terms in the current are analyzed across a broad range of model parameters. We propose that our theory can be tested experimentally by measuring the dependence of the anomalous Hall resistivity on the in-plane magnetic field in the surface states of the topological insulator SnTe, in WTe$_2$ and WSe$_2$ monolayers,  as well as, under applied uniaxial stress, in the Kondo lattice material
Ce$_3$Bi$_4$Pd$_3$ at very low temperatures. 
\end{abstract}

\maketitle


\section{Introduction}
Studies of the Hall effect have long held an important place in both experimental and theoretical condensed matter physics. In the conventional Hall effect, the transverse conductivity $\sigma_{xy}$ describes the system's response to an external electric field and arises from the orbital motion of conduction electrons coupled to an external magnetic field \cite{Abrikosov}. In contrast, the anomalous Hall effect originates either from the Zeeman coupling of electrons to an external magnetic field or, in the absence of a field, from a nonzero net magnetization.
In clean systems, the finite contribution to the anomalous Hall effect is determined by the Berry curvature of the Bloch wavefunctions, which modifies the electron velocity~\cite{AHE1995,AHE1999,Haldane2004,AboutHall1,AboutHall2,AboutHall3,Magnus1,Magnus2}. In both cases, however, it is evident that within linear response theory a nonzero Hall conductivity requires the breaking of time-reversal symmetry.

The next natural step in the study of the Hall effect has been to generalize the theory to nonlinear response~\cite{NHE2021,Nonlinear1,Nonlinear2,Nonlinear3,Nonlinear4,Nonlinear5,Konig2021}. In particular, it has been shown that a nonlinear Hall-like current can arise in systems lacking inversion symmetry, regardless of whether time-reversal symmetry is preserved~\cite{Spivak2009,Sodemann2015,Inti2}. This contribution to the conductivity is geometric in origin and is commonly referred to as the Berry curvature dipole (BCD) contribution~\cite{Sodemann2015,BerryDipole1,BerryDipole2,BerryDipole3,BerryDipole4,BerryDipole5,Zhang_2018,BCD1,BCD2,BCD3,BCD4,BCD6}. Most recently, it has also been theoretically proposed that the nonlinear Hall effect can be induced by superconducting fluctuations~\cite{Dong2025}.

During the past few years several papers appeared reporting on the experimental observation of the nonlinear Hall effect in time-reversal invariant electronic systems~\cite{NHEexp1,NHEexp2,NHEexp3,NHEexp4,NHEexp5,NHEexp6,NHEexp7}. Furthermore, giant nonlinear Hall effect has also been observed in an $f$-electron system Ce$_3$Bi$_4$Pd$_3$~\cite{Silke2021}. What makes this material special is the presence of strong electronic correlations, which, thanks to the partially filled $f$-electronic orbitals on Ce ions, drive the formation of the local magnetic moments. It has been argued that this material belongs to the class of Weyl-Kondo semimetals \cite{Silke2021,Weyl-Kondo}. Although it was suggested that BCD may be responsible for the magnitude of the nonlinear Hall response, the onset of the effect below $T^*\approx 3$K still remains not fully understood~\cite{DMFT2024,Setty2025} for the following reason.

\begin{figure}
\includegraphics[width=0.9750\linewidth]{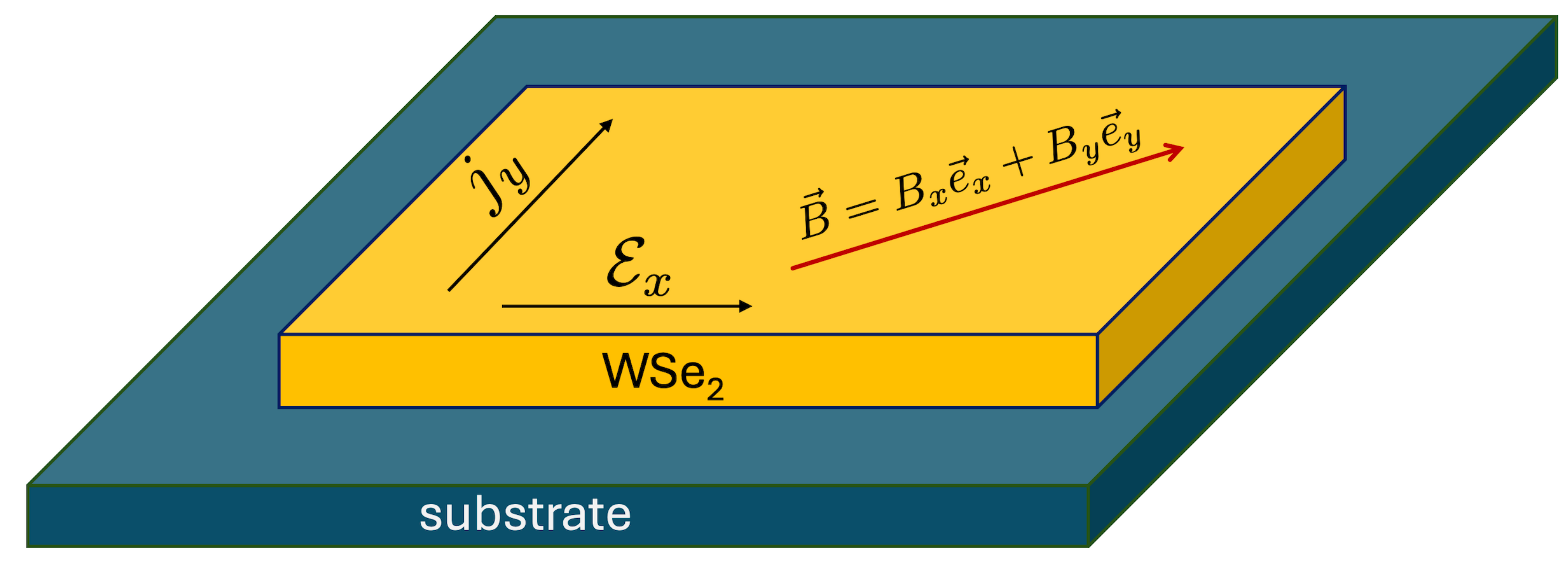}
\caption{Schematic representation of the experimental setup in which the electric field ${\cal E}_x$ is applied along the $x$-axis of the two-dimensional metallic film (e.g., a single layer of WSe$_2$) and electric current density $j_y$ is measured along the $y$-axis. The external magnetic field is applied in the plane of the film. The application of magnetic field induces the nonlinear Hall response, $j_y\propto {\cal E}_x^2$. Nontrivial geometric properties of the underlying band structure also contribute to the nonlinear Hall effect whose magnitude is determined by the Berry curvature dipole. Depending on the direction of magnetic field as well as value of the microscopic model parameters, these two contributions may either enhance or cancel each other.}
\label{Fig-Main}
\end{figure}

The intrinsic BCD contribution to the nonlinear Hall response requires the crystal point
group to be gyrotropic, i.e.\ to admit a nonvanishing second-rank axial tensor. Among the
gyrotropic classes the two cubic ones, $T$ and $O$, admit only the isotropic form
$D_{ab}\propto\delta_{ab}$, for which the nonlinear Hall current
$\propto\varepsilon_{adc}D_{bd}\mathcal{E}_b\mathcal{E}_c$ vanishes identically. A nonzero
BCD-driven Hall response therefore requires one of the remaining classes.
For example, Ce$_3$Bi$_4$Pd$_3$ crystallizes in $I\bar43d$ with point group $T_d$, which is not
gyrotropic: the three $C_2$ axes force all off-diagonal components of $D_{ab}$ to vanish,
$C_3$ along $[111]$ equates the diagonal ones, and any improper element --- a $\{110\}$
mirror or $S_4$ --- then sends $D\to-D$, so $D_{ab}\equiv0$. A nonzero BCD in this material
consequently requires the cubic symmetry to be lowered, for instance by uniaxial strain,
which reduces $T_d$ to $D_{2d}$, $C_{3v}$ or $C_{2v}$ for stress along $[001]$, $[111]$ or
$[110]$ respectively --- all gyrotropic. For $[001]$ stress the allowed form is
$D = \mathrm{diag}(D_1,-D_1,0)$. Since the dipole then vanishes linearly with the
symmetry-breaking parameter, exactly as we will  show below for the tilt $\alpha$, this yields a
sharp experimental handle: the zero-field second-harmonic Hall response of
Ce$_3$Bi$_4$Pd$_3$ should grow from zero linearly in applied uniaxial stress.

More generally, in two-dimensional systems lacking inversion symmetry, the nonlinear Hall response can be tuned by an external in-plane magnetic field, as illustrated in Fig.\ref{Fig-Main}~\cite{Edel1988,Magarill90,Dmitriev91}. Depending on its orientation, the field may either enhance or suppress the intrinsic contribution. The objective of the present work is to provide a detailed microscopic investigation of this effect.

In what follows, we consider a rather simple microscopic model of a two-dimensional disordered electronic system with the nontrivial geometric band structure which gives rise to both nonzero Berry curvature and BCD. Specifically, we will use the model of two tilted massive Dirac cones. To account for the effect of the in-plane magnetic field, we assume that the mass has a weak quadratic dispersion. 

Using the theory of nonequilibrium systems, we derive the quantum kinetic equation and compute the second-order response of the current density to an applied electric field in the presence of an in-plane magnetic field. Our approach enables the analysis of both the 
$ac$- and $dc$-limits. We further identify the conditions under which the magnetic-field contribution can become of the same order as the geometric Berry curvature dipole contribution.

Our paper is organized as follows. In the next section, we introduce the model and provide the details of the derivation of the quantum kinetic equation for the Wigner distribution function. In Section~\ref{Sec:Nlr}, we present the calculation of the nonlinear contribution to the current density. In the first part of Section~\ref{Sec:Nlr}, we analyze the nonlinear current in the absence of a magnetic field and re-derive the Berry curvature dipole contribution, while in the second part we examine the contribution arising from the external magnetic field. Section~\ref{Sec:Disc} is devoted to a discussion of our results and conclusions. Finally, Appendices~\ref{Linearsigmaxy}–\ref{app:model} contain the technical details of the transport calculations and model derivation. Throughout this work, we use the units with $\hbar =c = 1$.


\section{Model and Formalism}
In this Section, we describe the formalism and provide the list of main equations which are used to calculate the second harmonic of the current density. 

\subsection{Tilted massive Dirac fermion}

We begin by considering the following two-band model Hamiltonian in two spatial dimensions~\cite{Du2021,Dong2025}: 
\beg\label{Eq1}
\begin{split}
\hat{H}_{\pm K}(\bk)&=\pm \alpha k_x\hat{\sigma}_0\pm vk_x\hat{\sigma}_x+ vk_y\hat{\sigma}_y+m\hat{\sigma}_z.
\end{split}
\en
In crystalline topological insulators Pauli matrices $\hat\sigma_i$ act in a space of underlying Kramers doublets and $m$ is determined by the magnitude of the ferroelectric distortion~\cite{Liu2013,Serbyn2014}. 
Here we focus on the strained transition metal dichalcogenides (TMDs) realization of the Hamiltonian \eqref{Eq1} (see App.~\ref{app:model} for details).
In these systems the Pauli matrices act in the space of orbitals comprising the conduction and valence bands.
Based on DFT calculations the conduction band at $\pm \bm{K}$ are formed by the $d_{z^2}$ atomic transition metal orbitals.
The valence band at $\pm \bm{K}$ has a $d_{x^2-y^2} \pm i d_{xy}$ orbital character \cite{Fang2015,Model1}.
Both valence and conduction band wave functions are even in the basal plane mirror operation $(m_z)$.

\begin{figure}
\includegraphics[width=0.9750\linewidth]{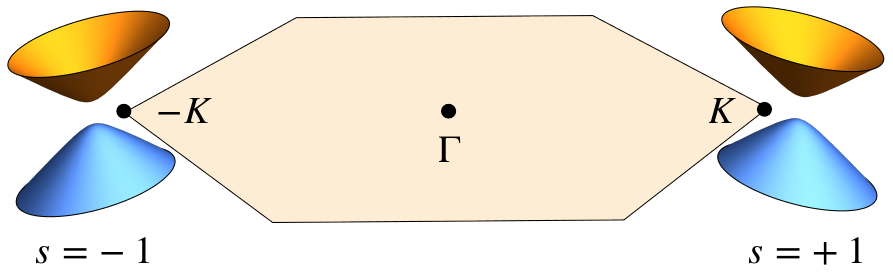}
\caption{Schematic low-energy band structure for monolayer WTe$_2$. The two massive tilted Dirac cones are located at momenta $\pm K$ in the hexagonal Brillouin zone, as described by Eq.~\eqref{Eq1}, and $s=\pm1$ is the valley index. }
\label{Fig:tilted}
\end{figure}

The model \eqref{Eq1} ignored the actual spin. 
In principle one has two copies of the Hamiltonian \eqref{Eq1} per spin. This Hamiltonian represents a minimal effective model for the surface states in crystalline topological insulator SnTe and
monolayer WTe$_2$ as it describes electrons with dispersion determined by the two tilted massive Dirac cones located at momenta $\pm K$ of the two-dimensional Brillouin zone,  as shown in Fig.~\ref{Fig:tilted} \cite{Model3,Model1,Model2}. 

Furthermore, we assume the scalar potential disorder $U(\br)$ with the white-noise Gaussian statistics $\left\langle U(\br)U(\br')\right\rangle_{\textrm{dis}}={\delta(\br-\br')}/({2\pi\nu_F\tau})$, where $\tau^{-1}$ is the disorder scattering rate and $\nu_F$ is the single-particle density of states per spin at the Fermi level. In the first part of this work, we will be interested in transport properties in the absence of a magnetic field, and for this reason it has not been included into Eq.~(\ref{Eq1}). Lastly, parameter $\alpha$ in Eq.~(\ref{Eq1})  accounts for the tilt of the Dirac cones.

We find it convenient to work in the eigenbasis of Hamiltonian~(\ref{Eq1}),
$\hat{H}_{s K}(\bk)=\sum_{\eta=1}^2\veps_{\bk s}^{(\eta)}|u_{s\eta}(\bk)\rangle\langle u_{s\eta}(\bk)|$, where  $s=\pm1$ is the valley index and the eigenenergies are given by 
\beg\label{Eigenvals}
\veps_{\bk s}^{(\eta)}=s\alpha k_x+(-1)^\eta\sqrt{m^2+v^2k^2}. 
\en
The corresponding eigenvectors are
\beg\label{Eigenvecs}
\begin{split}
|u_{s1}(\bk)\rangle&=\frac{1}{\sqrt{2}}\left(\begin{aligned} &-s\sqrt{1-\frac{m}{b_\bk}}e^{-is\varphi_\bk} \\
& \sqrt{1+\frac{m}{b_\bk}}
\end{aligned}\right), \\ 
|u_{s2}(\bk)\rangle&=\frac{1}{\sqrt{2}}\left(\begin{aligned} &s\sqrt{1+\frac{m}{b_\bk}}e^{-is\varphi_\bk} \\
& \sqrt{1-\frac{m}{b_\bk}}
\end{aligned}\right).
\end{split}
\en
Here we have introduced the shorthand notation $b_\bk=\sqrt{m^2+v^2k^2}$ and $\tan\varphi_\bk=k_y/k_x$. 

Next, we introduce the velocity operator which is defined according to
$\hat{v}_{a}^{(s)}(\bk)=\partial_{k_a}\hat{H}_{s}(\bk)$.
Denoting $\bn=\bk/k$, the matrix elements of the velocity operator $\hat{v}_x^{(s)}$ in the basis (\ref{Eigenvecs}) are given by
\beg\label{vxs}
\hat{v}_x^{(s)}=\left[\begin{matrix} s{\alpha}-\frac{v^2k_x}{b_\bk} & \frac{v}{k}\left(\frac{mk_x}{b_\bk}-isk_y\right) \\
\frac{v}{k}\left(\frac{mk_x}{b_\bk}+isk_y\right) & s{\alpha}+\frac{v^2k_x}{b_\bk}
\end{matrix}\right].
\en
Similarly, for the matrix elements of the velocity operator $\hat{v}_y^{(s)}$ we have
\beg\label{vys}
\hat{v}_y^{(s)}=\left[\begin{matrix} -\frac{v^2k_y}{b_\bk} & \frac{v}{k}\left(\frac{mk_y}{b_\bk}+isk_x\right) \\
\frac{v}{k}\left(\frac{mk_y}{b_\bk}-isk_x\right) & \frac{v^2k_y}{b_\bk}
\end{matrix}\right].
\en

It is well known that the band structure described by Eq.~(\ref{Eq1}) possesses nontrivial geometric phase. Introducing the Berry connection ${\vec {\mathbf {\cal A}}^{(\eta)}}(\bk)=i\langle \eta\bk|{\vec \nabla}_\bk|\eta \bk\rangle$ and the Berry curvature 
$\Omega_a^{(\eta)}(\bk)=\epsilon^{abc}\partial_b{\cal A}_c^{(\eta)}(\bk)$, we find that $\Omega_x^{(\eta)}=\Omega_y^{(
\eta
)}=0$ and 
\beg\label{BerryPhase}
\Omega_z^{(\eta)}(\bk)=\left(\frac{smv^2}{2b_\bk^3}\right)(-1)^{1+\eta}.
\en
As one may have expected, the Berry curvature is independent of the tilt parameter $\alpha$ and is proportional to the energy gap in the Dirac spectrum and the valley index. The latter turns out to be crucial for tje nonlinear Hall effect. 
\subsection{Quantum kinetic equation}
The key quantity in our subsequent analysis is the Wigner distribution function (WDF) which is defined according to 
\beg\label{WDF}
\begin{aligned}
w_{\alpha\beta}({\vec k},{\vec r};\eps,t)&=\frac{1}{2\pi}\int d^2{\vec \rho}\int d\eta e^{i{\vec k}{\vec \rho}-i\eps\eta}\\&\times\left\langle\psi_\beta\dg\left({\vec r}+\frac{{\vec \rho}}{2},t+\frac{\eta}{2}\right)\psi_\alpha\left({\vec r}-\frac{{\vec \rho}}{2},t-\frac{\eta}{2}\right)\right\rangle.
\end{aligned}
\en
Here $\psi_\alpha^\dagger(\br,t)$ and $\psi_\alpha(\br,t)$ are the usual creation and annihilation fermionic fields for a given valley and averaging is performed over the ground state of the Hamiltonian (\ref{Eq1}). Spatial homogeneity of WDF is restored after averaging over disorder. In the absence of external fields, we find for the WDF computed in the basis of Eq.~(\ref{Eigenvecs}):
\beg\label{BareWDF}
\hat{w}_{\bk\eps}^{(0)}=f(\eps)\left(
\begin{matrix} \delta\left(\eps-\eps_{\bk s}^{(1)}\right) & 0 \\
0 & \delta\left(\eps-\eps_{\bk s}^{(2)}\right)\end{matrix}
\right),
\en
where $f(\eps)$ is the Fermi-Dirac distribution function. In the presence of a spatially uniform electric field, the WDF satisfies the following quantum kinetic equation~\cite{Andrey2006,Dzero2024}
\beg\label{Eq4w}
\begin{aligned}
&\left(\frac{\partial}{\partial t}+\frac{1}{\tau}\right)\hat{w}_{\bk\eps}(t)+i\left\{{\mathbf b}_\bk\cdot{\mbox{\boldmath $\sigma$}},\hat{w}_{\bk \eps}(t)\right\}_{-}\\&=-
\alpha\tilde{\nabla}_x\hat{w}_{\bk \eps}(t)-\frac{v}{2}\left\{\hat{\mbox{\boldmath $\sigma$}},\tilde{\mbox{\boldmath $\nabla$}}\hat{w}_{\bk \eps}(t)\right\}_{+}.
\end{aligned}
\en
Here $\{\hat{f},\hat{g}\}_{\pm}=\hat{f}\hat{g}\pm\hat{g}\hat{f}$ and
\beg\label{tnabla}
\tilde{\mbox{\boldmath $\nabla$}}\hat{w}_{\bk\eps}(t)=\frac{e{\mathbf E}(t)}{\omega}
\left(\hat{w}_{\bk\eps+\frac{\omega}{2}}(t)-\hat{w}_{\bk\eps-\frac{\omega}{2}}(t)\right).
\en
In the kinetic equation (\ref{Eq4w}), we have also neglected the collision integral beyond $\tau$-approximation, which is technically equivalent to neglecting vertex corrections due to disorder scattering. In the limit when electric field is weak, kinetic equation (\ref{Eq4w}) can be solved by perturbation theory. For example, for the monochromatic electric field 
\beg\label{DefineEfield}
{\mathbf E}(t)={\vec {\cal E}}e^{-i\omega t}+\mathrm{c.c.}, 
\en
the first order correction to the WDF is 
\beg\label{FirstOrder}
\begin{aligned}
\left[\hat{w}_{\bk\eps}^{(1)}\right]_{aa}&=-\left(\frac{e}{\omega z_\omega}\right)\left([\hat{\mathbf v}]_{aa}\cdot{\vec{\cal E}}\right)
\left(\hat{w}_{\bk\eps+\frac{\omega}{2}}^{(0)}-\hat{w}_{\bk\eps-\frac{\omega}{2}}^{(0)}\right)_{aa}, \\
\left[\hat{w}_{\bk\eps}^{(1)}\right]_{a\overline{a}}&=-\left(\frac{eZ_{\bk\omega}^{(a)}}{2\omega}\right)\left([\hat{\mathbf v}]_{a\overline{a}}\cdot{\vec{\cal E}}\right)\sum\limits_{b=1}^2
\left(\hat{w}_{\bk\eps+\frac{\omega}{2}}^{(0)}-\hat{w}_{\bk\eps-\frac{\omega}{2}}^{(0)}\right)_{bb},
\end{aligned}
\en
where $a=(1,2)$, $\overline{1}=2$, $\overline{2}=1$, $z_\omega=-i\omega+\tau^{-1}$ and $Z_{\bk\omega}^{(a)}=[z_\omega+2ib_\bk\cos(\pi a)]^{-1}$.

\subsection{Linear response}
Given the Wigner distribution function, the current density can be computed as
\beg\label{current}
{\mathbf j}(t)=-e\int\frac{d^2\bk}{(2\pi)^2}\int\limits_{-\infty}^{\infty}\textrm{Tr}[\hat{\mathbf v}
\hat{w}_{\bk\eps}(t)]d\eps,
\en
where $\hat{v}_x$ and $\hat{v}_y$ are the $x$- and $y$-components of the velocity operator
with the matrix elements given by Eqs.~(\ref{vxs},\ref{vys}). One can check the correctness of this definition by computing the current density in the linear response approximation. Without loss of generality we can ignore the tilt by setting $\alpha=0$. Specifically, in the limit $\omega\ll \mu$ we have reproduced the Drude formula $\sigma_{xx}=e^2{D}\nu_F/(1-i\omega \tau)$ and also linear intrinsic contribution to the anomalous Hall effect $\sigma_{xy}=se^2/4\pi$ (see  Appendix~\ref{Linearsigmaxy} for details).


\section{Nonlinear response \label{Sec:Nlr}}

\subsection{Berry curvature dipole contribution}
Before analyzing the role of the in-plane magnetic field, we first demonstrate the microscopic derivation of the geometric contribution to the second harmonic using the method of quantum kinetic equation. This contribution is determined by the Berry curvature dipole, which in two dimensions is defined as~\cite{Sodemann2015} 
\begin{equation}  
{\cal D}_i \equiv
\sum\limits_{s=\pm}\int\frac{d^2\bk}{(2\pi)^2}\Omega_z(\bk) \frac{\partial f(\eps_{\bk s})}{\partial k_i},  \label{Eq:BCDdef}
\end{equation}
with $i = x, y$. We find that it is proportional to the tilt parameter $\alpha$, Eq.~(\ref{Eq1}).

Given our choice of the model, the Berry curvature dipole will be pointing along the $x$-axis (see Eq. \eqref{FirstBerry} below). We find it more convenient to consider the situation when electric field lies in the $xy$-plane and compute the current along the $x$-axis.  In this case the nonlinear Hall contribution to the current is defined by 
\beg\label{nlHall}
{j}_x^{(2\omega)}=\chi_{xxy}(\omega){\cal E}_x{\cal E}_y.
\en

An interesting property of the quantum kinetic equation~(\ref{Eq4w}) is that the second order corrections to the WDF are formally given by the same expressions as the first order ones, Eq.~(\ref{FirstOrder}), in which we replace $\hat{w}_{\bk\eps}^{(0)}\to\hat{w}_{\bk\eps}^{(1)}$. Straightforward analysis shows that the contribution to $\chi_{xxy}$ from the diagonal matrix elements of the velocity operator and WDF given by $[\hat{v}_x]_{11}
[\hat{w}_{\bk\eps}^{(2)}]_{11}+[\hat{v}_x]_{22}
[\hat{w}_{\bk\eps}^{(2)}]_{22}\propto s\alpha k_x^2$ which gives zero upon the summation over valleys, $s=\pm$. Below we will see that in order to get nonzero Hall response one will need to include an in-plane magnetic field. We therefore conclude that nonlinear correction to the current will necessarily originate from the scattering processes which involve the off-diagonal matrix elements of the WDF.  

We thus analyze the remaining contribution
$[\hat{v}_x]_{12}[\hat{w}_{\bk\eps}^{(2)}]_{21}+[\hat{v}_x]_{21}
[\hat{w}_{\bk\eps}^{(2)}]_{12}$. With the help of expressions for $[\hat{w}_{\bk\eps}^{(2)}]_{a\overline{a}}$ (see Appendix \ref{AppendixC} for details), we find that the combination of velocity matrix elements $[\hat{v}_x]_{12}[\hat{v}_x]_{21}[\hat{v}_y]_{aa}\propto k_y$ gives zero upon integration over $k_y$. Therefore, the only nonzero contribution will arise from the following combinations: $[\hat{v}_y]_{12}[\hat{v}_x]_{21}[\hat{v}_x]_{aa}$ and $[\hat{v}_x]_{12}[\hat{v}_y]_{21}[\hat{v}_x]_{aa}$. In particular, with the help of expressions (\ref{vxs},\ref{vys}) we find
\beg\label{vprods}
\begin{aligned}
&[\hat{v}_y]_{21}[\hat{v}_x]_{12}=-\frac{v^2}{b_\bk}\left({ism}+\frac{v^2k_xk_y}{b_\bk}\right), \\
&[\hat{v}_y]_{12}[\hat{v}_x]_{21}=\frac{v^2}{b_\bk}\left({ism}-\frac{v^2k_xk_y}{b_\bk}\right).
\end{aligned}
\en
Again keeping in mind the subsequent integrations over momentum components, one realizes that the terms $\propto k_xk_y$ do not contribute to the second harmonic since the resulting expression under the integral will depend only on $k_x$ and $k_x^2+k_y^2$. 
As a result, we derive the following expression for the response function $\chi_{xxy}$:
\beg\label{SecondHarmonic}
\begin{aligned}
\chi_{xxy}&=\left(\frac{2e^3}{\omega^2z_\omega}\right)\sum\limits_{s=\pm}\int\frac{d^2\bk}{(2\pi)^2}
\left(\frac{smv^3}{z_{2\omega}^2+4b_\bk^2}\right)\\&\times\int\limits_{-\infty}^{\infty}\left(\frac{vk_x}{b_\bk}\right){\cal W}({\bk\eps};\omega)d\eps,
\end{aligned}
\en
where function ${\cal W}({\bk\eps};\omega)$ is defined according to 
\beg\label{DefW}
\begin{aligned}
{\cal W}(\bk\eps;\omega)&=\sum\limits_{a=1}^2(-1)^a\left[\hat{w}_{\bk\eps+{\omega}}^{(0)}+\hat{w}_{\bk\eps-{\omega}}^{(0)}-2\hat{w}_{\bk\eps}^{(0)}\right]_{aa}
\end{aligned}. 
\en
Equation~\eqref{SecondHarmonic} is derived in the limit $\alpha\ll v$ and ignores contributions of order $O(\alpha^2)$.
\begin{figure}
\includegraphics[width=0.9250\linewidth]{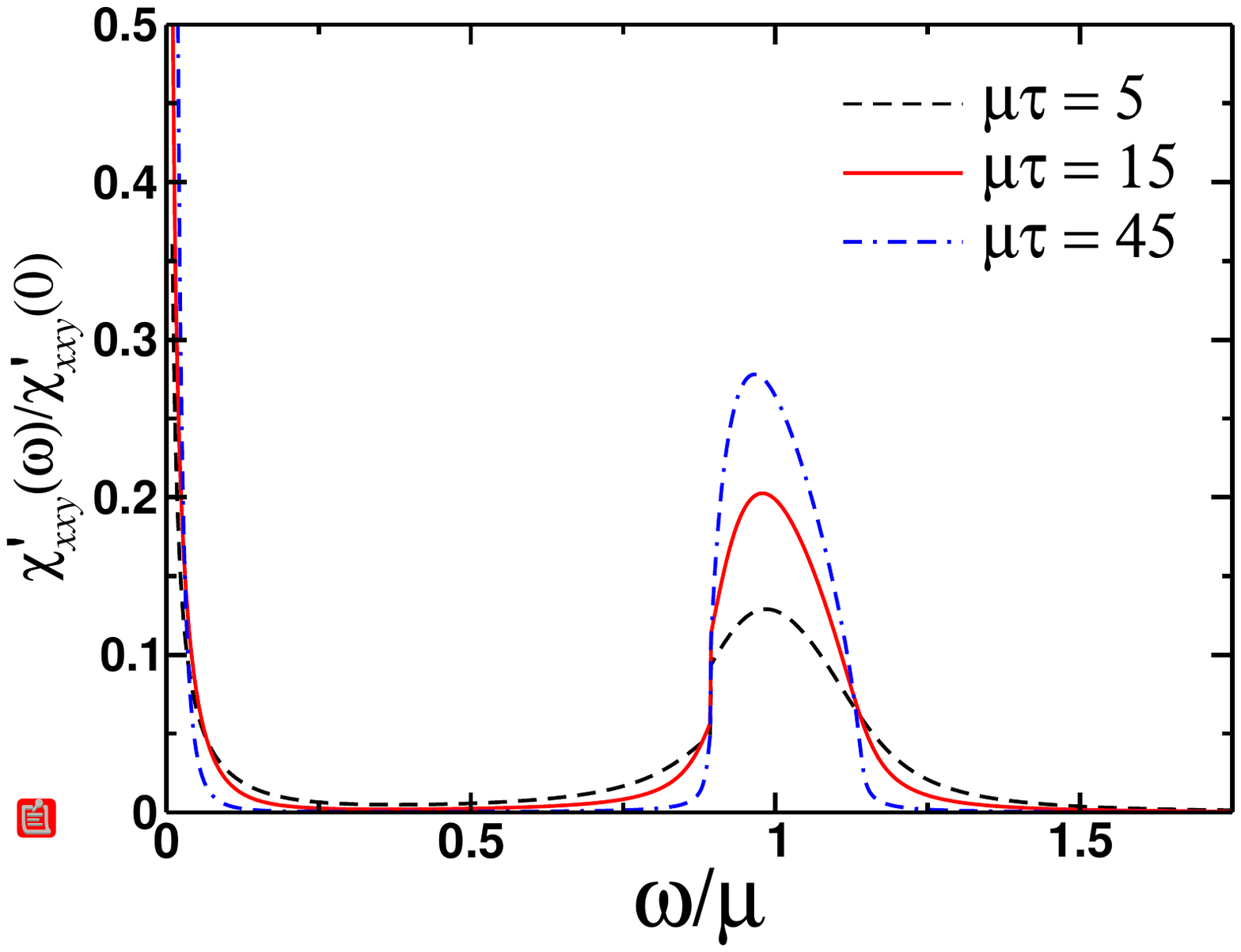}
\caption{Frequency dependence of the real part of the nonlinear response function $\chi_{xxy}=\chi_{xxy}'+i\chi_{xxy}''$ (\ref{SecondHarmonic}) for different values of the disorder scattering rate. In order to generate this plot we used $\alpha/v=0.125$ and $m/\mu=0.33$. }
\label{Fig-Resonance}
\end{figure}
In Fig.~\ref{Fig-Resonance}, we present the frequency dependence of the real part of the response function, $\chi_{xxy}(\omega)=\chi_{xxy}'(\omega)+i\chi_{xxy}''(\omega)$. Notably, function $\chi_{xxy}'(\omega)$ exhibits a peak at frequency $\omega\approx\mu$. This peak describes the resonant transitions between the valence and conduction band. The location of the peak is deduced from the fact that the denominator in the expression under the integral (\ref{SecondHarmonic}) approaches zero as $\tau\to\infty$. This resonant effect is not captured by the semiclassical Boltzmann equation approach~\cite{Sodemann2015} (see also \cite{Culcer2022} and references therein).

\paragraph{Low frequency limit.} We consider the limit of low frequencies first. Specifically, we take $\omega\ll m$ and for weak disorder $m\tau\gg 1$ we have $z_{2\omega}\ll 2b_\bk$, so that the expression under the integral (\ref{SecondHarmonic}) simplifies to
\beg\label{SecondHarmonicA}
\begin{aligned}
\chi_{xxy}&\approx\sum\limits_{s=\pm}\left(\frac{e^3v^4}{2\omega^2}\right)\left(\frac{sm}{z_{\omega}}\right)
\int\frac{d^2\bk}{(2\pi)^2}\frac{k_x}{b_\bk^3}\\&\times\left[f(\eps_{\bk s}+\omega)+
f(\eps_{\bk s}-\omega)-2f(\eps_{\bk s})\right].
\end{aligned}
\en
Here we assumed for definiteness that the chemical potential lies in the upper band, $\mu>0$.
Expanding the distribution functions in powers of $\omega$ yields
\beg\label{SecondHarmonicApp}
\begin{aligned}
\chi_{xxy}\approx\sum\limits_{s=\pm}\frac{sme^3v^2\tau}{2(1-i\omega\tau)}
\int\frac{d^2\bk}{(2\pi)^2}\frac{v^2k_x}{b_\bk^3}f''(\eps_{\bk s}).
\end{aligned}
\en
Next, we express function $\chi_{xxy}$ explicitly in terms of the integral over the Berry curvature. To do that we replace the expression under the integral using the following identity
\beg\label{AuxRel1}
\frac{v^2k_x}{b_\bk}f''(\eps_\bk)=\frac{\partial}{\partial k_x}f'(\eps_{\bk s})-s\alpha f''(\eps_{\bk s}).
\en
At this point it will be convenient to separate the contributions from different valleys, so we write $\chi_{xxy}=\sum\limits_{s=\pm}\chi_{xxy}^{(s)}$ with
\beg\label{Precious1}
\begin{aligned}
\chi_{xxy}^{(s)}&=\frac{sme^3v^2\tau}{2(1-i\omega\tau)}
\int\frac{d^2\bk}{(2\pi)^2}\frac{1}{b_\bk^2}
\frac{\partial}{\partial k_x}[f'(\eps_{\bk s})]\\&-\frac{m\alpha e^3v^2\tau}{2(1-i\omega\tau)}
\int\frac{d^2\bk}{(2\pi)^2}\frac{1}{b_\bk^2}f''(\eps_{\bk s}).
\end{aligned}
\en
Integrating by parts in the first term and using Eq.~(\ref{AuxRel1}) again we find:
\beg\label{Precious3}
\begin{aligned}
\chi_{xxy}^{(s)}&=\frac{e^3\tau}{(1-i\omega\tau)}
\int\frac{d^2\bk}{(2\pi)^2}\left(\frac{smv^2}{b_\bk^3}\right)\frac{\partial f(\eps_{\bk s})}{\partial k_x}
\\&-\frac{m\alpha e^3v^2\tau}{2(1-i\omega\tau)}
\int\frac{d^2\bk}{(2\pi)^2}\left\{\frac{f'(\eps_{\bk s})}{b_\bk^3}+\frac{f''(\eps_{\bk s})}{b_\bk^2}\right\}.
\end{aligned}
\en

Let us first consider the first term in Eq.~\eqref{Precious3}. Given the expressions for the Berry curvature, Eq.~(\ref{BerryPhase}), and the Berry curvature dipole, Eq.~\eqref{Eq:BCDdef}, one immediately recognizes that the first integral in Eq.~(\ref{Precious3}) is of  geometric origin: 
\beg\label{FirstBerry}
\chi_{xxy}^{(s)}=\frac{e^3\tau}{(1-i\omega\tau)}
\int\frac{d^2\bk}{(2\pi)^2}\left(\frac{smv^2}{b_\bk^3}\right)\frac{\partial f(\eps_{\bk s})}{\partial k_x}
\equiv\frac{e^3\tau{\cal D}_x^{(s)}}{1-i\omega\tau},
\en
where we calculated for the Berry curvature dipole per valley 
\beg\label{IntBerryD}
{\cal D}_x^{(s)}=\frac{3\alpha m|\mu|\tilde{\gamma}^2v^3\left(v^2-\alpha^2\right)}{4\pi(\mu^2v^2+m^2\alpha^2)^{5/2}}
\en
and we defined the parameter $\tilde{\gamma}=\sqrt{\mu^2v^2/(v^2-\alpha^2)-m^2}$. This contribution does not depend on the valley index $s = \pm1$, so that ${\cal D}_x^{(+)} = {\cal D}_x^{(-)} = {\cal D}_x/2$, where ${\cal D}_x$ is the total Berry curvature dipole.  After subsequent summation over valleys, we find $\chi_{xxy}=e^3\tau{\cal D}_x/(1-i\omega\tau)$. For details of the evaluation of the momentum integral in Eq.~(\ref{FirstBerry}) we refer the reader to Appendix~\ref{AppendixA}. 

As for the remaining integral in Eq.~(\ref{Precious3}), one notices that it is proportional to $\alpha$ and after subsequent rearrangement of the terms it will give the contribution $O(\alpha^2)$ which we ignore. In principle, it can be shown that this contribution will exactly cancel the corresponding contribution from the similar term in Eq.~\eqref{SecondHarmonic}, which we have ignored as well. This cancellation is not accidental and is guaranteed by the time-reversal symmetry of the system. 

Equation (\ref{IntBerryD}) represents the only nonzero component of the Berry curvature dipole. Indeed, as expected on purely symmetry grounds, we find that the second harmonic response is linearly proportional to the tilt parameter $\alpha$ which breaks the mirror symmetry and gives rise to nonzero Berry curvature dipole contribution ${\vec {\cal D}}={\cal D}_x{\vec e}_x$. 
Finally, in the dc limit , $\omega\tau\ll1$, and assuming also $v\gg \alpha$, $\mu\gg m$, we find for the response function after summation over the two valleys 
\beg\label{chixxyDx}
[\chi_{xxy}]_{\omega\tau\ll 1}\approx\left(\frac{3e^3}{2\pi}\right)\frac{m\alpha\tau}{\mu^2}.
\en

The calculation presented above can of course be repeated assuming the electric field is aligned along the $x$-axis, while the current is flowing along the $y$-axis, i.e., only the $y$-components of the velocity operator in Eq.~(\ref{current}) needs to be considered. As a result, the general expression for the nonlinear current can be written as 
\beg\label{j2wfin}
{\vec j}_{2\omega}=\frac{e^3\tau\left({\vec{\cal D}}\cdot{\vec{\cal E}}\right)}{1-i\omega\tau}\left({\vec{\cal E}}\times{\vec e}_z\right).
\en
Absolutely analogous expression can be derived for the dc second-order current, with $\left({\vec{\cal E}}\times{\vec e}_z\right)$ replaced by $\left({\vec{\cal E^*}}\times{\vec e}_z\right)$. {We note that a factor of 2 difference with Ref.~\cite{Sodemann2015} originates from a slightly different definition of electric field and current.}

\paragraph{High frequency limit.} We now turn our attention to the analysis of $\chi_{xxy}(\omega)$, Eq. (\ref{SecondHarmonic}), in the limit of high frequencies, $\omega \sim m$ and $\omega\tau\gg 1$ while keeping $\omega\ll \mu$. The latter condition still allows us to expand the arguments of the distribution functions up to the second order in powers of $\omega/\mu$. The remaining part of the calculation is very similar to that leading us to Eq.~(\ref{Precious3}). As a result, we find that at high frequencies 
\beg\label{chihigh}
[\chi_{xxy}]_{\omega\tau\gg 1}\sim\frac{im\alpha}{\omega\mu^2},
\en
which is purely dissipative. 

As the frequency increases, the second-order response exhibits a sharp resonance around $\omega\approx \mu$, as shown in Fig.~\ref{Fig-Resonance}. At even higher frequencies,  $\omega\gg \mu$, we find that $[\chi_{xxy}]_{\omega\tau\gg 1}\propto i/\omega^3$.

\subsection{Effect of in-plane magnetic field}
As we have discussed so far, Eq.~(\ref{j2wfin}) represents the only nonzero second-order Hall response without an external magnetic field ${\mathbf B}$. In order to elucidate the contribution from an in-plane magnetic field, we assume that it couples to spin $\hat {\boldsymbol \sigma}$ through the Zeeman coupling. We will discuss the following model Hamiltonian 
\beg\label{NewH}
\begin{aligned}
\hat{\cal H}_{\pm K}(\bk)&=\pm\alpha k_x\hat{\sigma}_0\pm v k_x\hat{\sigma}_x+vk_y\hat{\sigma}_y\\&+\gamma(B_x\hat{\sigma}_y\pm B_y\hat{\sigma}_x)+\left(m+\frac{\beta k^2}{2}\right)\hat{\sigma}_z
\end{aligned}
\en
which represents the extension of the model of the TMDs, Eq.~\eqref{Eq1},
derived in Appendix~\ref{app:model} based on the symmetry considerations applied to a TMD monolayer.
The mass dispersion in Eq.~\eqref{NewH}, $\beta k^2 \hat \sigma_z/2$, which is naturally present in realistic band structures, will turn out to be crucial for our analysis. 
On a substrate there is a finite admixture of the $m_z$ odd orbitals $d_{xz} $ and $d_{yz}$ that is necessary for a finite $\gamma$.
In principle, we could have also considered nonlinear-in-momentum corrections to the Dirac dispersion, however, while amenable to a numerical analysis, producing analytically tractable calculations for this case turns out to be quite challenging. 

It is instructive to emphasize the symmetry distinction between the two contributions to the nonlinear Hall response. The Berry-curvature-dipole term is geometric in origin and is allowed even in a time-reversal-symmetric system: although the Berry curvature $\Omega_z(\bk)$ is odd under time reversal and has opposite signs in the two valleys, the tilt of the Dirac cones renders the Fermi-surface average that defines the dipole ${\cal D}_x$ even under time reversal, so that this contribution survives the summation over valleys already at $B=0$.
The magnetic-field-induced contribution has an entirely different symmetry origin. As the semiclassical analysis of Appendix \ref{App:Boltzmann} makes clear, it is controlled solely by the band dispersion near the Dirac nodes (see the second term of Eq. (D1)), and in a time-reversal-symmetric system this term is odd under time reversal and cancels exactly between the two valleys. A finite in-plane Zeeman field breaks time-reversal symmetry and lifts this cancellation, thereby generating a nonzero dispersion-only contribution that is linear in B. In this sense the field-induced nonlinear Hall response is a direct manifestation of time-reversal-symmetry breaking, in contrast to the Berry-curvature-dipole response, which is permitted even when time-reversal symmetry is preserved.

Expressions for the basis functions can be easily generalized by noting that in the absence of the tilt and mass dispersion ($\alpha=\beta=0$) the effect of magnetic field results in the shift of momentum: 
\beg\label{k2q}
k_x\to q_x=k_x+\frac{\gamma}{v}B_y,  ~k_y\to q_y=k_y+\frac{\gamma}{v}B_x.
\en 
Primarily for computational convenience, we consider the setup in which the direction of the electric field points along the $x$-axis, ${\vec {\cal E}}={\cal E}{\vec e}_x$, and compute the second harmonic of the current along the $y$-axis $j_y^{(2\omega)}$ (see Fig.~\ref{Fig-Main}). Given that the Berry curvature dipole is pointing along the $x$-axis, this will also allow us to make a direct comparison with the BCD contribution to the current, Eq. \eqref{j2wfin}. 

The weakly dispersive mass term $\propto \beta k^2$ modifies the velocity operators according to
\beg\label{vxq}
\hat{v}_x^{(s)}=\left[\begin{matrix} s\alpha-\frac{v^2q_x}{\eps_\bq} -\frac{\beta m_\bk k_x}{\eps_\bq} & \frac{v}{q}\left(\frac{m_\bk q_x}{\eps_\bq}-\frac{\beta k_xq^2}{\eps_\bq}-isq_y\right) \\
\frac{v}{q}\left(\frac{m_\bk q_x}{\eps_\bq}-\frac{\beta k_xq^2}{\eps_\bq}+isq_y\right) & s\alpha+\frac{v^2q_x}{\eps_\bq}+\frac{\beta m_\bk k_x}{\eps_\bq}
\end{matrix}\right]
\en
and
\beg\label{vyq}
\hat{v}_y^{(s)}=\left[\begin{matrix} -\frac{v^2q_y}{\eps_\bq} -\frac{\beta m_\bk k_y}{\eps_\bq} & \frac{v}{q}\left(\frac{m_\bk q_y}{\eps_\bq}-\frac{\beta k_yq^2}{\eps_\bq}+isq_x\right) \\
\frac{v}{q}\left(\frac{m_\bk q_y}{\eps_\bq}-\frac{\beta k_yq^2}{\eps_\bq}-isq_x\right) & \frac{v^2q_y}{\eps_\bq}+\frac{\beta m_\bk k_y}{\eps_\bq}
\end{matrix}\right].
\en
Here we introduced $q=\sqrt{(k_x+\gamma B_y/v)^2+(k_y+\gamma B_x/v)^2}$, $m_\bk=m+\beta k^2/2$, and $\eps_\bq=(v^2q^2+m_\bk^2)^{1/2}$. 
In what follows, we will be focusing solely on contributions from the magnetic field and for this reason we will set $\alpha=0$. 
\paragraph{Second order corrections.} We proceed by looking at the second-order corrections in powers of electric field to the $y$-component of the current. As in the derivation of the Berry curvature dipole contribution to the nonlinear current, one should distinguish between the contributions from the diagonal parts of the WDF $[\hat{w}^{(2)}]_{aa}$ and the off-diagonal ones $[\hat{w}^{(2)}]_{\overline{a}a}$.  We show in Appendix~\ref{AppendixC} that the off-diagonal contribution has additional small factor $1/(\mu \tau)^2 \ll 1$ compared to the diagonal one, see Eq.~\eqref{joffSmallBeta}.

\begin{figure}
\includegraphics[width=0.9250\linewidth]{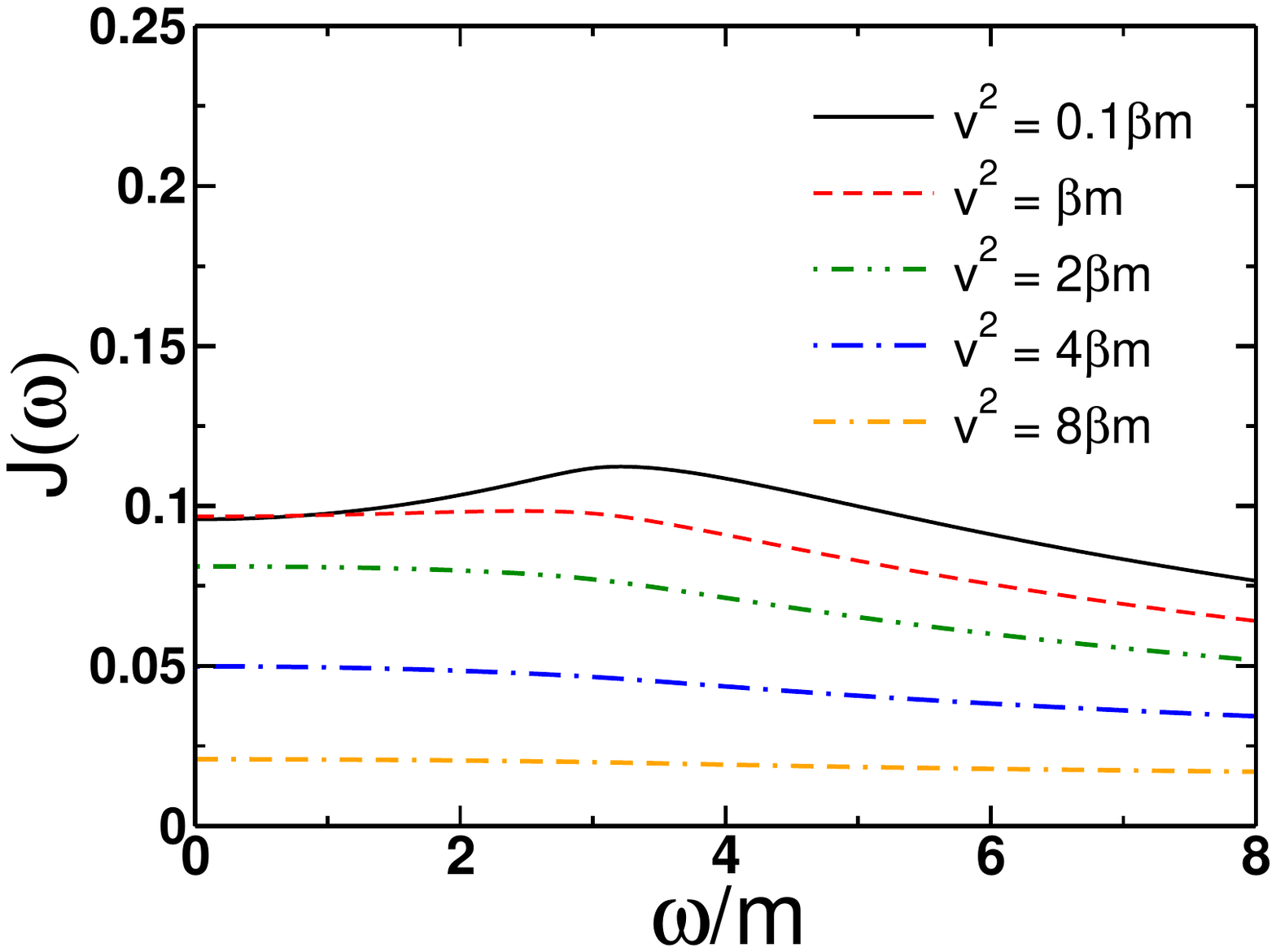}
\caption{Frequency dependence of the function $J(\omega)$, Eq. \eqref{Current2Plot}, shown here for different values of the dimensionless parameter $v^2/(\beta m)$. The value of the chemical potential has been fixed to $\mu=4m$.}
\label{Fig-Jw}
\end{figure}

Thus we focus on the contribution from the {\it diagonal} components of the WDF. We have:
\beg\label{Diags}
\begin{aligned}
\sum\limits_{\mathrm{a}=1}^2[\hat{v}_y^{(s)}]_{\mathrm{aa}}&\left[\hat{w}_{\bk\eps}^{(2)}\right]_{\mathrm{a}\mathrm{a}}\approx\left(\frac{2e^2{\cal E}^2}{\omega^2z_\omega z_{2\omega}}\right)\sum\limits_{\mathrm{a}=1}^2[\hat{v}_y^{(s)}]_{\mathrm{aa}}\left([\hat{v}_x^{(s)}]_{\mathrm{aa}}\right)^2\\&\times\left(2\left[\hat{w}_{\bq\eps}^{(0)}\right]_{\mathrm{aa}}-\left[\hat{w}_{\bq\eps+\omega}^{(0)}\right]_{\mathrm{aa}}-\left[\hat{w}_{\bq\eps-\omega}^{(0)}\right]_{\mathrm{aa}}\right).
\end{aligned}
\en
In this expression, we have used the fact that $[\hat{v}_y^{(s)}]_{11}+[\hat{v}_y^{(s)}]_{22}=0$, see Eq.~\eqref{vyq}. 
In the limit of small magnetic fields, we find to linear order in $\bf B$
\beg\label{Current2Plot}
j_{y}^{(2\omega)}({\mathbf B})=\frac{2\gamma B_xe^3{\cal E}^2\tau^2J(\omega)}{(1-i\omega\tau)(1-2i\omega\tau)}.
\en
{The dimensionless function $J(\omega)$ is given by Eq.~\eqref{SumOfJs} from  Appendix~\ref{AppendixC}, and its frequency dependence is shown in Fig.~\ref{Fig-Jw}. For frequencies $\omega\leq \mu-m$, the momentum integral for $J(\omega)$ can be evaluated analytically, resulting in Eq.~\eqref{ComputeDamnedIntegral}.}


At small frequencies, $\omega \ll \mu$, we obtain a compact expression:
\beg\label{MainResultjy2w}
\begin{aligned}  
j_y^{(2\omega)}({\mathbf B})&\approx\frac{e^3{\cal E}^2(\beta \tau k_F^2)^2}{(1-i\omega\tau)(1-2i\omega\tau)}\frac{\gamma v B_x}{2\pi \mu^3} \frac{\mu^2 - m^2_{k_F}+2m \, m_{k_F}}{\mu^2 + m^2_{k_F} - 2m\, m_{k_F}}. 
\end{aligned}
\en

To elucidate the physical origin of this result, we reproduce it (up to a factor of 2) using semiclassical Boltzmann approach in Appendix~\ref{App:Boltzmann}. We show that it is determined solely by the dispersion relation near the Dirac nodes and {\it is not} related to the geometry of the band structure. Indeed, in the absence of tilt, the Berry curvature of Hamiltonian~\eqref{NewH} has the opposite sign in two valleys even at finite magnetic field, hence the Berry curvature dipole contribution vanishes. However, this cancellation is accidental and is not enforced by any symmetry. We expect that the higher-order couplings to Zeeman field allowed by symmetry will result in finite geometrical contribution, and we leave this problem for future study.

Finally, we find that the longitudinal component of the current equals $j_{x}^{(2\omega)}({\mathbf B})=(3B_y/B_x)j_{y}^{(2\omega)}({\mathbf B})$. This relation constitutes one of the experimental predictions of our theory. 


\section{Discussion and conclusions \label{Sec:Disc}}

In addition to the intrinsic magnetic field dependent contribution calculated in this work, realistic materials also exhibit extrinsic mechanisms -- such as side-jump and scew-scattering processes -- that can contribute comparably to the nonlinear Hall response and are often independent of the disorder scattering rate~\cite{Golub2020,CIMOKE}. Although we did not include disorder-assisted effects in our analysis, we expect the predicted dependence of the nonlinear Hall effect on the magnetic field direction to remain robust and observable at sufficiently high fields. A detailed investigation of these extrinsic contributions is left for future work.


To estimate the magnitude of the effect studied in this work, we can use the following dimensionless parameter
\beg\label{eta}
\eta=\left(\frac{\gamma B_x}{m}\right)\left(\frac{\beta k_F^2}{\mu}\right)\left(\frac{\beta k_F}{\alpha}\right)\mu\tau
\en
that determines the ratio between the magnetic field and the Berry curvature dipole contributions to the nonlinear Hall response in the low-frequency limit $\omega\tau\ll 1$. In a disordered system ($m\tau\sim 1$) this parameter is quite small. However, in a relatively clean electronic system the smallness of the energy scale $\beta k_F^2$ can be partially compensated by the small value of $\alpha k_F$. For example, in WSe$_2$ and WTe$_2$  monolayers we can estimate $\mu\tau\approx 100$, $\beta k_F^2\sim 0.05\mu\approx 80$ meV and $\alpha k_F\approx 5$ meV~\cite{Culcer2022}, which yields $\eta\approx 10(\gamma B/m)$. For magnetic fields in the range $B\sim 1\div 10$ T, we find $\eta\sim 0.1\div 1$, which should be experimentally observable. The situation for Ce$_3$Bi$_4$Pd$_3$ seems even more favorable, as the electron's velocity is significantly reduced due to the strong hybridization between the conduction and $f$-electrons of the cerium ions.

Finally, we highlight the advantages of the quantum kinetic equation approach employed in this work. Unlike the semiclassical Boltzmann formalism, this method provides a unified framework that remains valid across a wide range of light frequencies, magnetic field strengths, and chemical potentials. It also naturally incorporates various forms of disorder through the collision integral, enabling a more realistic modeling of experimental conditions. Crucially, the quantum kinetic equation captures resonant interband transitions between the valence and conduction bands -- phenomena beyond the reach of Boltzmann-based treatments~\cite{Dzero2024}. Looking ahead, we plan to extend this method to explore orbital effects of strong magnetic fields in topological semimetals and flat-band systems~\cite{BCD5,BednikKozii2024,HassanzadeKozii2025}, as well as to investigate how generic disorder influences the quantization of the photogalvanic effect~\cite{deJuan2017,BCD8,Avdoshkin2020}.


In conclusion, we have explored the nonlinear Hall effect in two-dimensional Dirac materials under the influence of an external in-plane magnetic field. Using the quantum kinetic equation, we identified two distinct contributions to the second harmonic response. The first, of geometric origin, arises intrinsically from the Berry curvature dipole; the second is magnetic-field-induced and increases linearly with the field strength. Notably, the latter contribution exhibits a strong directional dependence, enabling the geometric response to be either enhanced or suppressed by adjusting the orientation of the applied field. Together, these findings suggest a practical and tunable tool for controlling nonlinear Hall phenomena in Dirac systems, offering clear experimental pathways for future investigation.

\section{Data availability}
The data that support the findings of this paper are not
publicly available upon publication because it is not technically feasible. Requests for further information or data should be sent to the authors.

\section{Acknowledgments}
We thank Leonid Golub for bringing several references to our attention and enlightening discussions. This work was financially supported by the National Science Foundation grant DMR-2400484 (M.D.) M. D. has performed this work in part at Aspen Center for Physics, which is supported by the National Science Foundation grant PHY-2210452. M.~K. acknowledges the support of the grant NSF-BSF DMR-2023693. V.~K. was supported by the Pittsburgh Quantum Institute Community Collaboration Award.

\begin{appendix}
\begin{widetext}
\section{Drude conductivity and anomalous Hall effect}\label{Linearsigmaxy}
It is useful to briefly discuss transport at the level of the linear response approximation. We start with the calculation of the longitudinal conductivity $\sigma_{xx}=j_x/{\cal E}_x$. Using the definitions \eqref{FirstOrder} and \eqref{current} in the main text, after a short calculation we find
\beg\label{jxDrude}
\sigma_{xx}=\frac{2e^2v^2\tau}{\omega(1-i\omega\tau)}\int\limits_{0}^\infty\frac{kdk}{2\pi}\left(\frac{z_\omega^2-2m^2+2b_\bk^2}{z_\omega^2+4b_\bk^2}\right)\left[\vartheta\left(\mu+\frac{\omega}{2}-b_\bk\right)-\vartheta\left(\mu-\frac{\omega}{2}-b_\bk\right)\right].
\en
and $b_\bk=\sqrt{m^2+(vk)^2}$. We note that this expression includes both contributions from diagonal as well as off-diagonal elements of $\hat{v}_\alpha$.
In deriving this formula we assumed the limit of $T\to 0$ and also that the chemical potential is in the conduction band. In the low-frequency limit $\omega\ll v k_F$ we recover the Drude formula
\beg\label{sigmaxxfin}
\sigma_{xx}=\frac{e^2}{\pi}\left(\frac{\mu\tau}{1-i\omega\tau}\right)\left[\frac{(1-i\omega\tau)^2+2(vk_F\tau)^2}{(1-i\omega\tau)^2+(2\mu\tau)^2}\right]\approx\left(\frac{e^2}{2\pi}\right)\frac1{1-i\omega \tau}\frac{\tau v^2k_F^2}{\mu}=\frac{e^2{D}\nu_F}{1-i\omega \tau}.
\en
Here ${D}=v_F^2\tau/2$ is the diffusion coefficient in two dimensions, $v_F=(\partial b_\bk/\partial k)_{k=k_F} = v^2 k_F/\mu$ is the Fermi velocity, $\nu_F=\mu/(\pi v^2)$ is the total density of states at the Fermi level and we assumed $\mu\tau\gg 1$.  

Let us now evaluate {\it linear} Hall conductivity $\sigma_{xy}=j_y/{\cal E}_x$. Again taking the limit $T\to 0$ we find
\beg\label{HallFin}
\sigma_{xy}^{(s)}={2sme^2v^2}\int\limits_{0}^\infty\frac{kdk}{2\pi}
\int\limits_0^{2\pi}\frac{d\phi}{2\pi}\sum\limits_{\eta=1}^2\frac{\delta\left(\mu-\eps_{\bk s}^{(\eta)}\right)}{z_\omega^2+4b_\bk^2},
\en
where $\mu$ is the chemical potential. Note that only off-diagonal components of the matrices $\hat{v}_x$ and $\hat{v}_y$ contribute to $\sigma_{xy}$. The prefactor of $2$ in front appears as a result of the partial cancellation of factor of $4$ which comes in combination with $b_\bk$ in the numerator and factor of $2$ in the denominator of $[\hat{w}^{(1)}]_{\textrm{a}\overline{\textrm{a}}}$, Eq. \eqref{FirstOrder}. We emphasize that  integral in Eq.~(\ref{HallFin}) can be rewritten as the integral over the Berry curvature, after performing the integration by parts.  
In the simplest case $\alpha=0$, the integral can be easily evaluated with the following result 
\beg\label{Hallalpha0}
[\sigma_{xy}]_{\alpha=0}
=\frac{se^2}{4\pi}\left[\frac{(2\mu\tau)^2}{(1-i\omega\tau)^2+(2\mu\tau)^2}\right]\frac{m}{\mu}.
\en
We consider the static limit $\omega\to 0$ in the absence of disorder, $\tau\to \infty$, but keeping $\omega\tau\ll 1$. Then, setting $\mu=m+0$ in Eq.~(\ref{Hallalpha0}), we readily recover the expression for the intrinsic contribution to the anomalous Hall conductivity for a single valley:
\beg\label{AHEint}
[\sigma_{xy}^{(\mathrm{int.})}]_{\alpha=0}=\frac{se^2}{4\pi}.
\en
The subsequent summation over the valley index leads to the cancellation of the corresponding contributions, as guaranteed by the time-reversal symmetry. The second-order Hall effect, which is the main focus of our work, survives even in time-reversal-symmetric systems.
\section{Berry curvature dipole contribution}\label{AppendixA}
In this Appendix, we provide the details of the calculation of the integral in Eq.~(\ref{FirstBerry}).
We consider the following integral:
\beg\label{JBD1}
J_{\textrm{B.d.}}=-\int\frac{d^2\bk}{(2\pi)^2}\frac{mv^2}{b_\bk^3}\frac{\partial f(\eps_\bk)}{\partial k_x}.
\en
In the low-temperature limit $T\to0$, the Fermi distribution function can be approximated by the step function 
\beg\label{T0}
f(\eps_\bk)\approx\vartheta(\mu-\eps_\bk).
\en
Thus, the integration in Eq.~(\ref{JBD1}) is performed over the surface
\beg\label{Surface}
\alpha k_x+\sqrt{m^2+v^2k^2}=\mu.
\en
We will integrate over $k_x$ and $k_y$ separately. We start with integrating over $k_y$ first by introducing a new variable
\beg\label{inty}
\begin{split}
&y=\sqrt{m^2+v^2k^2}.
\end{split}
\en
It follows that
\beg\label{Jbdint}
\begin{split}
J_{\textrm{B.d.}}=\frac{mv}{2\pi^2}\int\limits_{-\infty}^\infty
\frac{\left[\alpha(\mu-\alpha k_x)+{v^2k_x}\right]\vartheta(\mu-\alpha k_x-m_x)dk_x}{(\mu-\alpha k_x)^3\sqrt{(\mu-\alpha k_x)^2-m_x^2}},
\end{split}
\en
where $m_x=\sqrt{m^2+v^2k_x^2}$. We split Eq.~(\ref{Jbdint}) into two integrals:
\beg\label{JbggJbll}\nonumber
\begin{split}
J_{\textrm{B.d.}}^{(+)}&=\frac{mv}{2\pi^2}\int\limits_{0}^\infty\left(\alpha+\frac{v^2k_x}{\mu-\alpha k_x}\right)
\frac{\vartheta(\mu-\alpha k_x-m_x)dk_x}{(\mu-\alpha k_x)^2\sqrt{(\mu-\alpha k_x)^2-m_x^2}}, \\
J_{\textrm{B.d.}}^{(-)}&=\frac{mv}{2\pi^2}\int\limits_{0}^\infty\left(\alpha-\frac{v^2k_x}{\mu+\alpha k_x}\right)
\frac{\vartheta(\mu+\alpha k_x-m_x)dk_x}{(\mu+\alpha k_x)^2\sqrt{(\mu+\alpha k_x)^2-m_x^2}}.
\end{split} 
\en
Note that  $J_{\textrm{B.d.}}^{(+)}+J_{\textrm{B.d.}}^{(-)}=0$ when we set $\alpha=0$.
In order to evaluate the first integral, we change the integration variable according to
\beg\label{kx2qx}
k_x=q_x-\frac{\alpha\mu}{v^2-\alpha^2}.
\en
Given Eq.~(\ref{kx2qx}), it also follows that
\beg\label{AuxAux}
\begin{aligned}
&\alpha+\frac{v^2k_x}{\mu-\alpha k_x}=\frac{\alpha\mu+(v^2-\alpha^2)k_x}{\mu-\alpha k_x}=\frac{(v^2-\alpha^2)q_x}{\mu-\alpha k_x}, \\ 
&\mu-\alpha k_x=\frac{\mu v^2}{v^2-\alpha^2}-\alpha q_x\equiv \alpha (Q-q_x).
\end{aligned}
\en
Furthermore, we have
\beg\label{Root}
\begin{aligned}
(\mu-\alpha k_x)^2-m_x^2&=\frac{\mu^2v^2}{v^2-\alpha^2}-m^2-(v^2-\alpha^2)q_x^2\\&\equiv(v^2-\alpha^2)(\Lambda^2-q_x^2),
\end{aligned}
\en
where
\begin{equation}
\Lambda=\frac{1}{\sqrt{v^2-\alpha^2}}\sqrt{\frac{\mu^2v^2}{v^2-\alpha^2}-{m}^2}\equiv\frac{\gamma}{\sqrt{v^2-\alpha^2}}.
\end{equation}
Keeping in mind that $\alpha<v$, we note that 
\beg\label{QLambda}
Q=\frac{v}{\alpha}\left(\frac{\mu v}{v^2-\alpha^2}\right)> \Lambda.
\en
It will be convenient to use the shorthand notation
\beg\label{u}
u=\frac{\alpha}{\sqrt{v^2-\alpha^2}}.
\en
After we substitute these expressions into the integral and make a change of variables $x=q_x\sqrt{v^2-\alpha^2}$, we have
\begin{align}
J_{\textrm{B.d.}}^{(+)}(\alpha)&=\frac{mv(v^2-\alpha^2)^{3/2}}{2\pi^2\alpha^3}\int\limits_{u\mu}^\gamma
\frac{xdx}{\left[\left(\frac{v}{\alpha}\right)^2u\mu-x\right]^3\sqrt{\gamma^2-x^2}}\nonumber=
\frac{mv}{2\pi^2}\frac{(v^2-\alpha^2)^{3/2}}{\alpha^3\gamma^2}\int\limits_{u\mu/\gamma}^{1}\frac{qdq}{(a-q)^3\sqrt{1-q^2}} \label{Jbdp}
\end{align}
with $a=v^2 u \mu/\alpha^2 \gamma$. We note that 
\beg\label{Jpm}
\begin{split}
&J_{\textrm{B.d.}}^{(-)}(\alpha)=-J_{\textrm{B.d.}}^{(+)}(-\alpha)=-\frac{mv}{2\pi^2}\frac{(v^2-\alpha^2)^{3/2}}{\alpha^3\gamma^2}\int\limits_{-u\mu/\gamma}^{1}\frac{qdq}{(a+q)^3\sqrt{1-q^2}},
\end{split}
\en
which follows from the definition (\ref{JbggJbll}).  Changing under the integral $q\to -q$, we find
\beg\label{Sum}
\begin{aligned}
&J_{\textrm{B.d.}}(\alpha)=J_{\textrm{B.d.}}^{(+)}(\alpha)+J_{\textrm{B.d.}}^{(-)}(\alpha)=\frac{mv}{2\pi^2}\frac{(v^2-\alpha^2)^{3/2}}{\alpha^3\gamma^2}\int\limits_{-1}^{1}\frac{qdq}{(a-q)^3\sqrt{1-q^2}}=
\frac{mv}{2\pi^2}\frac{(v^2-\alpha^2)^{3/2}}{\alpha^3\gamma^2}\left[\frac{3\pi a}{2(a^2-1)^{5/2}}\right].
\end{aligned}
\en
This expression can be further simplified by using the following identities:
\beg\label{a2m1}
\begin{aligned}
a^2-1&=\frac{(\mu^2v^2+\alpha^2m^2)(v^2-\alpha^2)}{\alpha^2[(\mu^2-m^2)v^2+m^2\alpha^2]}, \quad
a=\frac{\mu v^2}{\alpha\sqrt{(\mu^2-m^2)v^2+m^2\alpha^2}}.
\end{aligned}
\en 
It then follows that
\beg\label{FinRes}
J_{\textrm{B.d.}}(\alpha)=
\left(\frac{\gamma^2}{4\pi}\right)\frac{3m\alpha\mu v^3(v^2-\alpha^2)}{(\mu^2v^2+m^2\alpha^2)^{5/2}}.
\en
{We emphasize that this expression should be viewed as the leading-order contribution to the second harmonic at small $\alpha$. Taking the limit $\alpha\ll v$ and $m\ll\mu$, we find}
\beg\label{FinResApp}
[J_{\textrm{B.d.}}]_{m\ll\mu}^{\alpha\ll v}\approx\frac{3}{4\pi}\left(\frac{m\alpha}{\mu^2}\right),
\en  
where we took into account that $[\gamma]_{m\ll\mu}\approx \mu$. 
\section{Corrections to the Wigner distribution function and nonlinear current in the presence of magnetic field}\label{AppendixC}
In the presence of magnetic field, the right hand side of the kinetic equation (\ref{Eq4w}) acquires an additional term 
$-\beta\left\{\bk\hat{\sigma}_z,\tilde{\mbox{\boldmath $\nabla$}}\hat{w}_{\bk \eps}(t)\right\}_{+}/2$. Because of this term, the corresponding expressions for the first-order corrections to the Wigner distribution function can be compactly expressed in terms of the matrix elements of the velocity operators as follows:
\beg\label{Eqs4w1B}
\begin{aligned}
&\left[\hat{w}_{\bq\eps}^{(1)}\right]_{aa}=-\left(\frac{e}{\omega z_\omega}\right)\left([\hat{\mathbf v}]_{aa}\cdot{\vec{\cal E}}\right)
\left[\hat{w}_{\bq\eps+\frac{\omega}{2}}^{(0)}-\hat{w}_{\bq\eps-\frac{\omega}{2}}^{(0)}\right]_{aa}, \quad
\left[\hat{w}_{\bq\eps}^{(1)}\right]_{a\overline{a}}=-\left(\frac{e}{2\omega}\right)\frac{\left([\hat{\mathbf v}]_{a\overline{a}}\cdot{\vec{\cal E}}\right)}{(z_\omega\pm 2i\eps_\bq)}
\sum\limits_{a=1}^2\left[\hat{w}_{\bq\eps+\frac{\omega}{2}}^{(0)}-\hat{w}_{\bq\eps-\frac{\omega}{2}}^{(0)}\right]_{aa}.
\end{aligned}
\en
Here the matrix elements of the velocity operators are given by Eqs.~(\ref{vxq})-(\ref{vyq}) of the main text. 

The second-order corrections to the WDF can be found similarly to Eq.~(\ref{Eqs4w1B}). For the correction to the first diagonal element of the WDF, we have
\beg\label{Eq4w211}
\begin{split}
\left[\hat{w}_{\bq\eps}^{(2)}\right]_{aa}=&-\left(\frac{e}{\omega}\right)\frac{\left([\hat{\mathbf v}]_{aa}\cdot{\vec{\cal E}}\right)}{z_{2\omega}}
\left[\hat{w}_{\bq\eps+\frac{\omega}{2}}^{(1)}-\hat{w}_{\bq\eps-\frac{\omega}{2}}^{(1)}\right]_{aa}-\left(\frac{e}{2\omega}\right)\frac{\left([\hat{\mathbf v}]_{a\overline{a}}\cdot{\vec{\cal E}}\right)}{z_{2\omega}}
\left[\hat{w}_{\bq\eps+\frac{\omega}{2}}^{(1)}-\hat{w}_{\bq\eps-\frac{\omega}{2}}^{(1)}\right]_{\overline{a}a}
\\&-\left(\frac{e}{2\omega}\right)\frac{\left([\hat{\mathbf v}]_{\overline{a}a}\cdot{\vec{\cal E}}\right)}{z_{2\omega}}
\left[\hat{w}_{\bq\eps+\frac{\omega}{2}}^{(1)}-\hat{w}_{\bq\eps-\frac{\omega}{2}}^{(1)}\right]_{a\overline{a}}.
\end{split}
\en
Here, $z_{2\omega}=(1-2i\omega\tau)/\tau$. 
Similarly, for the off-diagonal elements we find
\beg\label{w212}
\begin{split}
\left[\hat{w}_{\bk\eps}^{(2)}\right]_{12}=&
-\left(\frac{e}{2\omega}\right)\left(\frac{[\hat{\mathbf v}]_{12}\cdot{\vec{\cal E}}}{z_{2\omega}-2i\eps_\bq}\right)
\sum\limits_{a=1}^2\left[\hat{w}_{\bq\eps+\frac{\omega}{2}}^{(1)}-\hat{w}_{\bq\eps-\frac{\omega}{2}}^{(1)}\right]_{aa}, \quad
\left[\hat{w}_{\bk\eps}^{(2)}\right]_{21}=
-\left(\frac{e}{2\omega}\right)\left(\frac{[\hat{\mathbf v}]_{21}\cdot{\vec{\cal E}}}{z_{2\omega}+2i\eps_\bq}\right)
\sum\limits_{a=1}^2\left[\hat{w}_{\bq\eps+\frac{\omega}{2}}^{(1)}-\hat{w}_{\bq\eps-\frac{\omega}{2}}^{(1)}\right]_{aa}.
\end{split}
\en

We can employ these formulas to evaluate the Hall current. Specifically, let us compute the contributions which contain the diagonal elements of the velocity operators and consistent with our discussion in the main text we take the electric field to point along the $x$-axis, ${\vec{\cal E}}={\cal E}{\vec e}_x$.

We start with the contribution to the $y$-component of the current from the diagonal parts of the WDF:
\beg\label{DiagCurrent}
[j_y^{(2\omega)}]_{\mathrm{diag.}}=-{e}\sum\limits_{s=\pm}\int\frac{d^2\bq}{(2\pi)^2}\int\limits_{-\infty}^{\infty}\left([v_y^{(s)}]_{11}[\hat{w}_{\bq\eps}^{(2)}]_{11}+[v_y^{(s)}]_{22}[\hat{w}_{\bq\eps}^{(2)}]_{22}\right)d\eps
\en
Using \eqref{Eqs4w1B}, \eqref{Eq4w211} and \eqref{w212} we have
\beg\label{w211App}
\left[\hat{w}_{\bq\eps}^{(2)}\right]_{11}=\left(\frac{e^2{\cal E}^2}{\omega^2z_\omega z_{2\omega}}\right)[v_x^{(s)}]_{11}[v_x^{(s)}]_{11}\left[\hat{w}_{\bq\eps+\omega}^{(0)}+\hat{w}_{\bq\eps-\omega}^{(0)}-2\hat{w}_{\bq\eps}^{(0)}\right]_{11}+\left[\delta\hat{w}_{\bq\eps}^{(2)}\right]_{11}, 
\en
where for convenience we introduced 
\beg\label{deltaw11}
\left[\delta\hat{w}_{\bq\eps}^{(2)}\right]_{11}=\left(\frac{e^2}{2\omega^2}\right)\left(\frac{1-i\omega \tau}{1-2i\omega\tau}\right)\frac{[v_x^{(s)}]_{12}[v_x^{(s)}]_{21}{\cal E}^2}{(z_\omega^2+4\eps_\bq^2)}\left[\hat{w}_{\bq\eps+\omega}^{(0)}+\hat{w}_{\bq\eps-\omega}^{(0)}-2\hat{w}_{\bq\eps}^{(0)}\right]_{22},
\en
Similarly, we find
\beg\label{w222App}
\left[\hat{w}_{\bq\eps}^{(2)}\right]_{22}=\left[\delta\hat{w}_{\bq\eps}^{(2)}\right]_{11}+\left(\frac{e^2{\cal E}^2}{\omega^2z_\omega z_{2\omega}}\right)[v_x^{(s)}]_{22}[v_x^{(s)}]_{22}\left[\hat{w}_{\bq\eps+\omega}^{(0)}+\hat{w}_{\bq\eps-\omega}^{(0)}-2\hat{w}_{\bq\eps}^{(0)}\right]_{22},
\en
Given that matrix $\hat{v}_y$ is traceless, Eq. \eqref{vyq}, it follows
\beg\label{Sumdiag}
[v_y^{(s)}]_{11}[\hat{w}_{\bq\eps}^{(2)}]_{11}+[v_y^{(s)}]_{22}[\hat{w}_{\bq\eps}^{(2)}]_{22}=\left(\frac{e^2{\cal E}^2}{\omega^2z_\omega z_{2\omega}}\right)\sum\limits_{{\mathrm a}=1}^2{[v_y^{(s)}]_{{\mathrm a}{\mathrm a}}[v_x^{(s)}]_{{\mathrm a}{\mathrm a}}[v_x^{(s)}]_{{\mathrm a}{\mathrm a}}}\left[\hat{w}_{\bq\eps+\omega}^{(0)}+\hat{w}_{\bq\eps-\omega}^{(0)}-2\hat{w}_{\bq\eps}^{(0)}\right]_{{\mathrm a}{\mathrm a}}.
\en
Inserting these expressions into \eqref{DiagCurrent} and assuming that chemical potential lies in the conduction band $({\mathrm a}=2)$ and assuming that the frequenies do not exceed $\mu-m$ we then have
\beg\label{DiagCurrentFin}
[j_y^{(2\omega)}]_{\mathrm{diag.}}=\frac{e^3{\cal E}^2\tau^2}{\omega^2(1-i\omega\tau)(1-2i\omega\tau)}\sum\limits_{s=\pm}\int\frac{d^2\bq}{(2\pi)^2}\int  {[v_y^{(s)}]_{22}[v_x^{(s)}]_{22}[v_x^{(s)}]_{22}}\left[2\hat{w}_{\bq\eps}^{(0)}-\hat{w}_{\bq\eps+\omega}^{(0)}-\hat{w}_{\bq\eps-\omega}^{(0)}\right]_{22}d\eps.
\en
In order to keep the subsequent expressions compact we will ignore the terms which are proportional to the $y$-component of the magnetic field since, as one can easily check, the current will be proportional to $B_x$. We have ($\bq=\bk+{\mathbf h}$):
\beg\label{ProductAgain}
\begin{aligned}
[\hat{v}_y]_{22}[\hat{v}_x]_{22}[\hat{v}_x]_{22}&=\beta^3(u+m_\bk)^3\frac{k_yk_x^2}{\eps_{\bk+{\mathbf h}}^3}+\left(\frac{\beta}{\eps_\bk}\right)^3uh_y
(u+m_\bk)^2k_x^2\approx\frac{\beta^3(u+m_\bk)^3}{\eps_\bk^3}\left\{k_yk_x^2+uh_y\left[1-(u+m_\bk)\frac{3\beta k_y^2}{\eps_\bk^2}\right]k_x^2\right\},
\end{aligned}
\en
where $u=v^2/\beta$, $h_y=(\gamma/v)B_x$, $\eps_\bk=\sqrt{(vk)^2+m_\bk^2}$ 
and used the expansion
\beg\label{ExpandAgain}
\begin{aligned}
&\frac{1}{\eps_{\bk+{\mathbf h}}^3}\approx\frac{1}{\eps_\bk^3}-\frac{3v^2(\bk{\mathbf h})}{\eps_\bk^5}+O(h^2).
\end{aligned}
\en
We now have to insert \eqref{ProductAgain} into expression for the current. An analytic formula for the current can be obtained by changing the integration over momentum to the integration over energy $\xi=\sqrt{v^2k^2+m_\bk^2}$. In particular, we have
\beg\label{NewIntVar}
kdk=\frac{1}{\beta}\frac{\xi d\xi}{(m_\xi+u)}, \quad m_\xi=\sqrt{\xi^2+u\left(u+2m\right)}-u
\en
with $k^2=2(m_\xi-m)/\beta$.

The final result for the current can be expressed as a sum of three separate contributions. The first one originates from the expansion of the arguments of the distribution functions in powers of ${\mathbf h}$:
\beg\label{FirstInt}
J_a=\left(-\frac{\gamma v B_x}{4\pi}\right)\int\limits_m^\infty{(m_\xi+u)^2(m_\xi-m)^2}\left[2\delta(\mu-\xi)-\delta(\mu+\omega-\xi)-\delta(\mu-\omega-\xi)\right]\frac{d\xi}{\xi^3}.
\en
For our discussion below it proves convenient to represent the delta functions as the derivatives of the Heaviside step functions and perform the integration by parts. It obtains:
\beg\label{JaFin}
\begin{aligned}
J_a=\left(-\frac{\gamma v B_x}{4\pi}\right)\left(~\int\limits_{\Omega}^\mu-\int\limits_{\mu}^{\mu+\omega}~\right) \left[\frac{(-3)}{\xi^4}{(m_\xi+u)^2(m_\xi-m)^2}+\frac{2(m_\xi+u)(m_\xi-m)}{\xi^2}+\frac{2(m_\xi-m)^2}{\xi^2}\right]d\xi,
\end{aligned}
\en
where $\Omega=\mathrm{max}(m,\mu-\omega)$.

The second integral is defined by the second term in \eqref{ProductAgain}:
\beg\label{Jb}
J_b=\left(\frac{\gamma v B_x}{4\pi }\right)\left(~\int\limits_{\Omega}^\mu-\int\limits_{\mu}^{\mu+\omega}~\right)\frac{2(m_\xi+u)(m_\xi-m)}{\xi^2}{d\xi}.
\en
Lastly, the third integral will be given by 
\beg\label{Jc}
J_c=\left(\frac{\gamma v B_x}{4\pi}\right)\left(~\int\limits_{\Omega}^\mu-\int\limits_{\mu}^{\mu+\omega}~\right)\frac{(-3)}{\xi^4}{(m_\xi+u)^2(m_\xi-m)^2}{d\xi}.
\en
Comparing \eqref{Jb} and \eqref{Jc} with \eqref{JaFin}, we see that the sum of the first two terms in \eqref{JaFin} equals to $-J_b-J_c$. Introducing $J=J_a+J_b+J_c$  we have
\beg\label{SumOfJs}
J(\omega)=\left(-\frac{\gamma v B_x}{2\pi}\right)\left(~\int\limits_{\Omega}^\mu-\int\limits_{\mu}^{\mu+\omega}~\right)(m_\xi-m)^2\frac{d\xi}{\xi^2}.
\en
This integral can be computed analytically. Assuming that external frequency satisfies $\omega\le\mu-m$ so that  $\Omega=\mu-\omega$ we have
\beg\label{ComputeDamnedIntegral}
\begin{aligned}
J(\omega)&=\left(-\frac{\gamma v B_x}{2\pi}\right)\frac{2u}{\overline{\mu}{(\overline{\mu}^2-\overline{\omega}^2)}}\left\{[2-2r_\mu+\overline{m}(4+\overline{m}-2r_\mu)]\overline{\omega}^2+(1+\overline{m})\overline{\mu}\left[\overline{\mu}(2r_\mu-r_{\mu-\omega}-r_{\mu+\omega})+\overline{\omega}(r_{\mu+\omega}-r_{\mu-\omega})\right.\right.\\&\left.\left.(\overline{\mu}^2-\overline{m}^2)(f_{\omega+\mu}+f_{\omega-\mu}-2f_{\omega})\right]\right\}
\end{aligned}
\en
where we introduced general notation $\overline{a}=a/u$, $r_{\mu}=\sqrt{\mu^2+2m u+u^2}/u$ and $f_\mu=\sinh^{-1}(\mu/\sqrt{u^2+2mu})$, and we remind that the parameter $u$ has been defined as $u=v^2/\beta$. Expression \eqref{ComputeDamnedIntegral} can be further simplified in the limit of low frequencies $\omega\ll \mu$. As a result, for the current in this limit we find :
\beg
\begin{aligned}  \label{DamnedIntegralSmallw}
[j_y^{(2\omega)}]_{\mathrm{diag.}}&\approx\frac{e^3{\cal E}^2(\beta \tau k_F^2)^2}{(1-i\omega\tau)(1-2i\omega\tau)}\frac{\gamma v B_x}{2\pi \mu^3} \frac{\mu^2 - m^2_{k_F}+2m \, m_{k_F}}{\mu^2 + m^2_{k_F} - 2m\, m_{k_F}}. 
\end{aligned}
\en
In the limit $\beta \ll \min\{ v^2/m, v/k_F \}$, Eq.~\eqref{DamnedIntegralSmallw} can be further simplified to
\beg\label{JFin}
[j_y^{(2\omega)}]_{\mathrm{diag.}}\approx\frac{e^3{\cal E}^2\beta^2\tau^2}{(1-i\omega\tau)(1-2i\omega\tau)}\left(\frac{\mu^4-m^4}{2\pi v^2\mu^3}\right)\left(\frac{\gamma B_x}{v}\right)\approx\frac{e^3{\cal E}^2\beta^2\tau^2k_F^2}{(1-i\omega\tau)(1-2i\omega\tau)}\left(\frac{\mu^2+m^2}{2\pi \mu^3}\right)\left(\frac{\gamma B_x}{v}\right).
\en

Let us now discuss the contribution from the off-diagonal elements of the WDF. We will use the same approximations which lead us to \eqref{JFin}. By definition we have 
\beg\label{OffDiagCurrent}
[j_y^{(2\omega)}]_{\mathrm{off-diag.}}=-{e}\sum\limits_{s=\pm}\int\frac{d^2\bq}{(2\pi)^2}\int\limits_{-\infty}^{\infty}\left([v_y^{(s)}]_{12}[\hat{w}_{\bq\eps}^{(2)}]_{21}+[v_y^{(s)}]_{21}[\hat{w}_{\bq\eps}^{(2)}]_{12}\right)d\eps.
\en
We would like to remind the reader that off-diagonal parts of the WDF are the ones which gave rise to the Berry dipole contribution to the current. So, let us consider the combination under the integral in more details. Using \eqref{Eqs4w1B} along with \eqref{vxq} and \eqref{vyq} we find
\beg\label{Details}
\begin{aligned}
&[v_y^{(s)}]_{12}[\hat{w}_{\bq\eps}^{(2)}]_{21}+[v_y^{(s)}]_{21}[\hat{w}_{\bq\eps}^{(2)}]_{12}=\left(\frac{e{\cal E}v^2}{2\omega\eps_\bq^2}\right)\sum\limits_{a=1}^2\left[\hat{w}_{\bq\eps+\frac{\omega}{2}}^{(1)}-\hat{w}_{\bq\eps-\frac{\omega}{2}}^{(1)}\right]_{aa}\\&\times\left[\frac{v^2q_xq_y+\beta m_\bk(k_xq_y+k_yq_x)-\beta^2q^2 k_xk_y-isq_x\eps_\bq(m_\bk q_x-\beta k_xq^2)
-isq_y\eps_\bq(m_\bk q_y-\beta k_yq^2)}{z_{2\omega}+2i\eps_\bq}+\mathrm{c.c.}\right]
\end{aligned}
\en
It is clear that the terms which are linear in $s$ will give zero upon the summation over the valley index since we took $\alpha=0$. As a result the expression under the integral \eqref{OffDiagCurrent} simplifies to 
\beg\label{SimplifyUnderIntegral}
\begin{aligned}
&[v_y^{(s)}]_{12}[\hat{w}_{\bq\eps}^{(2)}]_{21}+[v_y^{(s)}]_{21}[\hat{w}_{\bq\eps}^{(2)}]_{21}=\left(\frac{e{\cal E}z_{2\omega}v^2}{\omega\eps_{\bq}^2}\right)\left(\frac{v^2q_xq_y-(\beta q)^2k_xk_y+\beta m_\bk(k_xq_y+k_yq_x)}{z_{2\omega}^2+4\eps_{\bq}^2}\right)\sum\limits_{a=1}^2\left[\hat{w}_{\bq\eps+\frac{\omega}{2}}^{(1)}-\hat{w}_{\bq\eps-\frac{\omega}{2}}^{(1)}\right]_{aa}\\&=\frac{(e{\cal E})^2\left({v^2+\beta m_\bk}\right)(vk_x)^2}{(1-i\omega\tau)(1-2i\omega\tau)}\left(\frac{1-2i\omega\tau}{\omega}\right)^2\left(\frac{(v^2+\beta m_\bk)q_y+\beta k_y(m_\bk-\beta q^2)}{(z_{2\omega}^2+4\eps_{\bq}^2)\eps_{\bq}^3}\right)\left[2\hat{w}_{\bq\eps}^{(0)}-\hat{w}_{\bq\eps+\omega}^{(0)}-\hat{w}_{\bq\eps-\omega}^{(0)}\right]_{22}.
\end{aligned}
\en
On the last step we took into account the definition of $[\hat{v}_x]_{22}$, used simple power counting to check that upon carrying out the momentum integration in \eqref{OffDiagCurrent} the result will be proportional to $h_y$ and we set $q_x=k_x$. 
Thus, expression \eqref{OffDiagCurrent} takes the following form
\beg\label{OffDiagCurrentSign}
[j_y^{(2\omega)}]_{\mathrm{off-diag.}}=-\frac{e^3{\cal E}^2(1-2i\omega\tau)}{\omega^2(1-i\omega\tau)}\sum\limits_{s=\pm}\int\left({v^2+\beta m_\bk}\right)(vk_x)^2\left(\frac{(v^2+\beta m_\bk)q_y+\beta k_y(m_\bk-\beta q^2)}{(z_{2\omega}^2+4\eps_{\bq}^2)\eps_{\bq}^3}\right)F_\bq(\omega)\frac{d^2\bq}{(2\pi)^2}
\en
and we introduced the shorthand notation
\beg\label{MyNotation}
F_\bq(\omega)=2\vartheta(\mu-\eps_\bq)-\vartheta(\mu+\omega-\eps_\bq)-\vartheta(\mu-\omega-\eps_\bq).
\en
Note that current from the off-diagonal components of the WDF \eqref{OffDiagCurrentSign} enters with the overall negative sign compared to \eqref{DiagCurrentFin}.

It remains to compute the linear-in-magnetic field correction to $[j_y^{(2\omega)}]_{\mathrm{off-diag.}}$. We use the following expansion
\beg\label{OffDiag1stExpand}
\frac{1}{(z_{2\omega}^2+4\eps_{\bq}^2)\eps_{\bq}^3}\approx\frac{1}{(z_{2\omega}^2+4\eps_{\bk}^2)\eps_{\bk}^3}-\frac{v^2k_yh_y(3z_{2\omega}^2+20\eps_\bk^2)}{(z_{2\omega}^2+4\eps_\bk^2)^2\eps_\bk^5}+O(h_y^2).
\en
and insert it into into \eqref{OffDiagCurrentSign}. In order to evaluate the integrals it will be useful to consider three different contributions. The first one originates from the second term in \eqref{OffDiag1stExpand}. After performing the integral over the direction of $\bk$ and $\eps$ we have
\beg\label{Joff1}
I_{1}=\left(-\frac{\gamma vB_x}{4\pi }\right)u\left(u+2m\right)\left(~\int\limits_{\mu-\omega}^\mu-\int\limits_{\mu}^{\mu+\omega}~\right)\frac{(m_\xi-m)^2(3z_{2\omega}^2+20\xi^2)d\xi}{(z_{2\omega}^2+4\xi^2)^2\xi^4}.
\en
Here $u=v^2/\beta$, we assumed for simplicity $\mu-\omega\geq m$, made a change of variables from $k$ to $\xi=\sqrt{v^2k^2+m_\bk^2}$ and used $\langle k_x^2k_y^2\rangle_{\hat{\bk}}=k^4/8$. Similarly, the second integral is defined as follows
\beg\label{Joff2}
I_{2}=\left(\frac{\gamma vB_x}{2\pi }\right)\left(u+m\right)\left(~\int\limits_{\mu-\omega}^\mu-\int\limits_{\mu}^{\mu+\omega}~\right)\frac{(m_\xi-m)d\xi}{(z_{2\omega}^2+4\xi^2)\xi^2}.
\en
Finally, the third integral that we need to consider appears as a result of expanding the distribution functions in powers of $h_y$:
\beg\label{Joff3}
I_{3}=\left(-\frac{\gamma vB_x}{4\pi }\right)u\left(u+2m\right)\int\limits_{m}^\infty\frac{(m_\xi-m)^2d\xi}{(z_{2\omega}^2+4\xi^2)\xi^3}[2\delta(\mu-\xi)-\delta(\mu+\omega-\xi)-\delta(\mu-\omega-\xi)].
\en
Next we write $\delta(\mu-\xi)=(-d/d\xi)\vartheta(\mu-\xi)$ and integrate by parts. This yields
\beg\label{Joff3ByParts}
\begin{aligned}
I_{3}=\left(\frac{\gamma vB_x}{4\pi }\right)u\left(u+2m\right)
\left(~\int\limits_{\mu-\omega}^\mu-\int\limits_{\mu}^{\mu+\omega}~\right)
\frac{(m_\xi-m)^2(3z_{2\omega}^2+20\xi^2)d\xi}{(z_{2\omega}^2+4\xi^2)^2\xi^4}-\left(\frac{\gamma vB_x}{2\pi }\right)\left(~\int\limits_{\mu-\omega}^\mu-\int\limits_{\mu}^{\mu+\omega}~\right)\frac{u\left(u+2m\right)(m_\xi-m)d\xi}{(m_\xi+u)(z_{2\omega}^2+4\xi^2)\xi^2}.
\end{aligned}
\en
The first term on the right hand side equals $-I_1$. Furthermore, in the second term we replace $u(u+2m)=(m_\xi+u)^2-\xi^2$. It then follows:
\beg\label{Sum123}
\begin{aligned}
I=I_1+I_2+I_3=\left(\frac{\gamma vB_x}{2\pi }\right)\left(~\int\limits_{\mu-\omega}^\mu-\int\limits_{\mu}^{\mu+\omega}~\right)\left(\frac{m_\xi-m}{m_\xi+u}-\frac{(m_\xi-m)^2}{\xi^2}\right)\frac{d\xi}{z_{2\omega}^2+4\xi^2}.
\end{aligned}
\en
Taking into account that realistically $\mu\tau\gg 1$ we can approximate
\beg\label{ZnamApprox}
\left(1-2i\omega\tau\right)^2+4(\mu\tau)^2\approx4(\mu\tau)^2.
\en
Since the resulting expression still remains quite cumbersome, we expand it in powers of $\beta$. This yields the following expression for $[j_y^{(2\omega)}]_{\mathrm{off-diag.}}$:
\beg\label{joffSmallBeta}
[j_y^{(2\omega)}]_{\mathrm{off-diag.}}\approx \frac{e^3{\cal E}^2m^2}{4\pi\mu^5}\left(\frac{\gamma B_x}{v }\right)\left(\frac{1-2i\omega\tau}{1-i\omega\tau}\right)\left(\frac{\beta m}{v}\right)^2\left(2-\frac{m^2}{\mu^2}\right).
\en
We emphasize that this result is smaller than the diagonal contribution, Eq.~\eqref{JFin}, by a large factor $(\mu \tau)^2 \gg 1$, and hence should be neglected to the leading order.


\section{Semiclassical Boltzmann approach \label{App:Boltzmann}}

It is instructive to discuss how Eq.~\eqref{JFin} can be obtained from the semiclassical Boltzmann theory. With the electric field along the $x$-axis, ${\vec {\cal E}}={\cal E}{\vec e}_x$, the corresponding expression for the second harmonic of the current density is given by~\cite{Sodemann2015}
\beg\label{SF1}
\bj^{(2\omega)}=\frac{e^3 \tau {\cal E}^2}{1-i\omega \tau}\vec e_y\int\frac{d^2\bk}{(2\pi)^2} \Omega_z(\bk)  \left[\frac{\partial}{\partial k_x}f_0(\eps_\bq)\right]   -\frac{e^3\tau^2{\cal E}^2}{(1-i\omega\tau)(1-2i\omega\tau)}
\int\frac{d^2\bk}{(2\pi)^2}\frac{\partial \eps_{\bq}}{\partial \bk}\left[\frac{\partial^2}{\partial k_x^2}f_0(\eps_\bq)\right].
\en
Here, $f_0(\eps)$ is the Fermi-Dirac distribution function and $\Omega_z(\bk)$ is the Berry curvature. Neglecting the tilt, $\alpha = 0$, and assuming the chemical potential is in the conduction band, the dispersion of Hamiltonian~\eqref{NewH} is given by $\eps_\bq = \sqrt{v^2 q^2 + m_\bk^2}$, with   $m_\bk = m + \beta k^2/2$ and $q$ given by Eq.~\eqref{k2q}. Note that the dispersion is identical for both valleys in the absence of tilt.

The first term in Eq.~\eqref{SF1} describes the Berry curvature dipole contribution. It is straightforward to show that the Berry curvature of Hamiltonian~\eqref{NewH} is strictly opposite in different valleys, even in presence of a finite magnetic field $\bB$. Combined with the symmetric dispersion at $\alpha = 0$, it implies that the Berry curvature dipole vanishes in this case. This cancellation is accidental and is not dictated by any symmetry. We expect the corresponding contribution to be nonzero once other symmetry-allowed terms with higher powers of $\bB$ are included in Eq.~\eqref{NewH}.

The second term in Eq.~\eqref{SF1} is determined solely by the band dispersion and vanishes in time-reversal-symmetric systems. Since finite Zeeman field in Eq.~\eqref{NewH} breaks time reversal, this term produces a nonzero linear-in-$B$ contribution to the second harmonic. The contributions from the two valleys in this case merely add up resulting in an extra factor of 2. 

The evaluation of the integral is most easily done in the zero-temperature limit, such that $f'_0(\eps) = - \delta(\eps - \mu)$. After straightforward manipulations with the derivatives of the delta-functions, we find to the linear order in $\bB$ (and quadratic in $\cal E$):
\beg  
\bj^{(2\omega)}({\mathbf B})\approx\frac{e^3{\cal E}^2(\beta \tau)^2}{ (1-i\omega\tau)(1-2i\omega\tau)}\frac{v k_F^4}{4\pi\mu^3}\frac{\mu^2 - m^2_{k_F}+2m \, m_{k_F}}{\mu^2 + m^2_{k_F} - 2m\, m_{k_F}} \left(\begin{array}{c} 3\gamma B_y \\ \gamma B_x  \end{array}  \right). 
\en
This expression reproduces Eq.~\eqref{MainResultjy2w}, up to a difference in factor of 2 which we cannot explain. We again stress that this result is derived in the absence of tilt, $\alpha = 0$. We anticipate that a nonzero $\alpha$ will change the leading power of $\beta$ in the final expression.

To make the connection between the Boltzmann and quantum kinetic approaches, we introduce 
\beg\label{calJ}
{\cal J}_{yxx}=\int\frac{d^2\bk}{(2\pi)^2}\frac{\partial \eps_{\bq}}{\partial k_y}\left[\frac{\partial^2}{\partial k_x^2}f_0(\eps_\bq)\right].
\en
Integrating by parts with respect to $k_x$ and then with respect to $k_y$ one can show that ${\cal J}_{yxx}={\cal J}_{xxy}$ where
\beg\label{calJxxY}
{\cal J}_{xxy}=\int\frac{d^2\bk}{(2\pi)^2}\frac{\partial \eps_{\bq}}{\partial k_x}\left[\frac{\partial^2}{\partial k_x\partial k_y}f_0(\eps_\bq)\right].
\en
Next, we rewrite the derivative of the distribution function in 
Eq.~\eqref{calJ} into the following form:
\beg\label{MassageDeriv}
\frac{\partial^2f_0}{\partial k_x^2}=\frac{\partial}{\partial k_x}\left[\frac{\partial \eps_\bq}{\partial k_x}\frac{\partial f_0}{\partial \eps_\bq}\right]=\left(\frac{\partial \eps_\bq}{\partial k_x}\right)^2\frac{\partial^2 f_0}{\partial \eps_\bq^2}+\frac{\partial^2 \eps_\bq}{\partial k_x^2}\left(\frac{\partial f_0}{\partial \eps_\bq}\right).
\en
The first term in this expression can be rewritten as 
\beg\label{LastTerm}
\left(\frac{\partial \eps_\bq}{\partial k_x}\right)^2\frac{\partial^2 f_0}{\partial \eps_\bq^2}=\left[\hat{v}_x^{(s)}\right]_{22}\left[\hat{v}_x^{(s)}\right]_{22}\cdot\lim\limits_{\omega\to 0}\left\{\frac{1}{\omega^2}\left[f_0(\eps_\bq+\omega)+f_0(\eps_\bq-\omega)-2f_0(\eps_\bq)\right]\right\},
\en
where $[\hat{v}_x^{(s)}]_{22}$, for a given valley $s$, has been defined in Eq.~\eqref{vxq}. On the other hand, the second term leads to 
\beg\label{SecondMassage}
\int\frac{d^2\bk}{(2\pi)^2}\frac{\partial \eps_{\bq}}{\partial k_y}\frac{\partial^2 \eps_\bq}{\partial k_x^2}\left(\frac{\partial f_0}{\partial \eps_\bq}\right)=\int\frac{d^2\bk}{(2\pi)^2}\frac{\partial^2 \eps_\bq}{\partial k_x^2}\left(\frac{\partial f_0}{\partial k_y}\right)
=-\int\frac{d^2\bk}{(2\pi)^2}\frac{\partial \eps_\bq}{\partial k_x}\left(\frac{\partial^2 f_0}{\partial k_x\partial k_y}\right)=-{\cal J}_{xxy}=-{\cal J}_{yxx}.
\en
After inserting Eq.~\eqref{MassageDeriv} into Eq.~\eqref{calJ} and using Eq.~\eqref{SecondMassage}, we obtain
\beg\label{calJAfter}
\begin{aligned}
{\cal J}_{yxx}&=\sum\limits_{s=\pm}\int\frac{d^2\bk}{(2\pi)^2}\left[\hat{v}_y^{(s)}\right]_{22}\left[\hat{v}_x^{(s)}\right]_{22}\left[\hat{v}_x^{(s)}\right]_{22}\frac{\partial^2 f_0}{\partial \eps_\bq^2}-{\cal J}_{yxx},
\end{aligned}
\en
which leads to
\beg\label{Finally}
{\cal J}_{yxx}=\lim\limits_{\omega\to 0}\frac{1}{2\omega^2}\sum\limits_{s=\pm}\int\frac{d^2\bk}{(2\pi)^2}\left[\hat{v}_y^{(s)}\right]_{22}\left[\hat{v}_x^{(s)}\right]_{22}\left[\hat{v}_x^{(s)}\right]_{22}\left[f_0(\eps_\bq+\omega)+f_0(\eps_\bq-\omega)-2f_0(\eps_\bq)\right],
\en
where integration now is over the vicinity of the corresponding valley instead of the whole Brillouin zone. Inserting this result back to Eq.~\eqref{SF1}, we recover the corresponding expression for the $y$-component of the current in Eq.~\eqref{DiagCurrentFin} obtained within the quantum kinetic approach, up to a factor of~$2$.  The reason for this discrepancy remains unclear.

\section{Derivation of the model Hamiltonian}
\label{app:model}
In this appendix we derive the Hamiltonian used in the present paper to compute the BCD and in-plane field contributions using the $\bm{k}\cdot \bm{p}$ approximation \cite{Liu2013a,Mockli2018}.
We start with the monolayer TMD ignoring the substrate.
In this case the point group symmetry of the lattice is $D_{3h}$ which we represent in the following form, 
$D_{3h} = C_{3v} \times \{ E, m_z \}$ where $E$ is the identity, $m_{x,y,z}$ represent the mirror in the plane that is orthogonal to $\hat{x}$, $\hat{y}$ and $\hat{z}$, respectively. 
And the $C_{3v}$ is the symmetry of the ammonia molecule. 
It is convenient to view the $C_{3v}$ group as generated by the three fold rotation, $C_3$ and the $m_x$ mirror.
We also ignore the spin for the moment.
The group of $\bm{K}$ is $C_3$ such that the Bloch states are the eigenstates of the three-fold rotation operation. 

The transition metal $d$-orbitals are split into two groups. 
The orbitals $\{ d_{z^2}, d_{x^2 - y^2}, d_{xy}\}$ or even in $m_z$ and $\{ d_{xz}, d_{yz}\}$ are odd. 
Hence following the DFT results we focus on the first set of three orbitals even in $m_z$.
At $\bm{K}$ the conduction band electrons has a $d_{z^2}$ character, and the valence band has a character $d_{x^2 -y^2} - i d_{xy}$ with the respective eigenvalues $1$ and $ \exp( i 4 \pi /3)$ as $C_3$ operation eigenvalues.
Here $r = \exp( i 2 \pi /3)$.
Following we write the effective Hamiltonian operating in the space formed by the conduction and valence bands,
$\Psi^{\pm \bm{K}}_{\bm{k}} = (\Phi^c_{\pm \bm{K}+\bm{k}},\Phi^v_{\pm \bm{K}+\bm{k}})^{tr}$, where $tr$ stands for the transposition and 
\begin{align}\label{k_dot_p13}
     \Phi^c_{\pm \bm{K}+\bm{k}}(\bm{r}) & =  \frac{1}{\sqrt{N}}\sum_{\bm{R}} e^{i (\bm{k} \pm \bm{K}) \bm{R}} d_{z^2}(\bm{r} - \bm{R})
     \notag \\
     \Phi^v_{\pm \bm{K}+\bm{k}}(\bm{r}) & =  \frac{1}{\sqrt{N}}\sum_{\bm{R}} e^{i (\bm{k} \pm\bm{K}) \bm{R}} d_{\mp}(\bm{r} - \bm{R})\, ,
\end{align}
where the summation is over the locations of the $N$ transition metal ions, and the shorthand notation 
$d_{\pm}(\bm{r}) = d_{x^2 -y^2}(\bm{r}) \pm i d_{xy}(\bm{r})$ has been introduced.
The basis at $\pm \bm{K}$ is not unique, yet our choice, \eqref{k_dot_p13} is convenient because of the simple relationship between the two basis sets
\begin{align}\label{choice}
    m_x \Phi^{c,v}_{\bm{K}} = \mathcal{T}  \Phi^{c,v}_{\bm{K}} = \Phi^{c,v}_{-\bm{K}}.
\end{align}
The effective two by two Hamiltonian at $\pm \bm{K}$ is by construction,
\begin{align}\label{Heff}
    \left[ H_{s {K}}(\bm{k}) \right]_{\alpha \beta} = \langle \Psi^{s \bm{K}}_{\bm{k},\alpha}| \mathcal{H} |  \Psi^{s\bm{K}}_{\bm{k},\beta}\rangle\, ,
\end{align}
where $\mathcal{H}$ is the full Bloch Hamiltonian.
We are employing the generic $\bm{k}\cdot \bm{p}$ method of getting $H_{\pm {K}}(\bm{k})$ which amounts to the series expansion of Eq.~\eqref{Heff} around $\pm \bm{K}$ constrained by the symmetry of the problem. 

We start with the constrains imposed by the group of $\bm{K}$ which contains just three elements and isomorphic $C_3$ group.
We therefore focus on one, say $\bm{K}$ valley.
And we only have to make sure that the Hamiltonian is invariant under the three fold rotation, $C_3$.
It is convenient to split the Hamiltonian \eqref{Heff} into the diaogonal and off-diagonal parts, $H^{d}$ and $H^o$.

The diagonal part contains the zero strain part, $H^{d_0}$ and the contribution of the strain, $H^{str}$.
At zero strain, it is insensitive to the orbital content, and therefore to the second order in momentum reads,
\begin{align}\label{diag_H}
    H^{d_0}_{s{K}}(\bm{k}) = \left(h_0 + h k^2\right)\sigma_0 + \left(m + \frac{\beta k^2}{2} \right)\sigma_z\, .
\end{align}
In Eq.~\eqref{diag_H} we only keep the second term as the first one is inconsequential. 
To add the effect of the strain we classify the strain components by the group of $\bm{K}$ ($C_3$),
\begin{align}
    C_3 (u_{xx} - u_{yy} \pm i 2 u_{xy}) = e^{\mp 4 i \pi /3} (u_{xx} - u_{yy} \pm i 2 u_{xy})\, .
\end{align}
Therefore, the invariants allowed by the group of $\bm{K}$ are real and imaginary parts of 
$k_+ (u_{xx} - u_{yy} + i 2 u_{xy}) $, and the preliminary Hamiltonian, 
\begin{align}
    H_{+\bm{K}}^{str} = A[ k_x (u_{xx} - u_{yy}) - 2 k_y u_{xy}] + B [ 2 k_x u_{xy} + k_y (u_{xx}-u_{yy})]\, .
\end{align}
As before, the $m_x$ symmetry gives 
\begin{align}
    H_{-\bm{K}}^{str} = -A[ k_x (u_{xx} - u_{yy}) - 2 k_y u_{xy}] - B [ 2 k_x u_{xy} + k_y (u_{xx}-u_{yy})]\, .
\end{align}
and the time reversal gives, 
\begin{align}
    H_{-\bm{K}}^{str} = -A[ k_x (u_{xx} - u_{yy}) - 2 k_y u_{xy}] + B [ 2 k_x u_{xy} + k_y (u_{xx}-u_{yy})]\, .
\end{align}
The consistency of the last two equations enforces, $B=0$, and in summary,
\begin{align}\label{strain19}
    H_{s\bm{K}}^{str} = s A[ k_x (u_{xx} - u_{yy}) - 2 k_y u_{xy}]\, . 
\end{align}
Since we are interested in the physical conditions with $m_x$ preserved, the strain we consider is such that $u_{xy}=0$,
therefore we get,
\begin{align}\label{strain21}
    H_{s\bm{K}}^{str} = s \alpha  k_x \, .
\end{align}
Let us note that the tensor $B_{i} B_j - (1/2) \bm{B}^2$ transforms in the same way as the components of the traceless strain tensor considered above.
Therefore, on par with Eq.~\eqref{strain19} we also have the contribution of the second order in magnetic field, 
\begin{align}\label{Bsquared}
    H_{s\bm{K}}^{d_0} = s A'[ k_x (B_x^2 - B_y^2) - 2 k_y B_x B_y]\, , 
\end{align}
which has a meaning of the two-fold modulation of the Dirac cone tilt as the in-plane magnetic field rotates.
Such two fold variation is normally associated with the planar Hall effect \cite{Goldberg1954,Zhong2023,Attias2024}.
The field induced dependent  tilt of the Dirac cones has been previously identified as the source of the linear planar Hall effect on a surface of the topological insulator \cite{Zheng2020}.
In these works, however the planar Hall has been induced by the terms in the Hamiltonian of the form, $B_x k_y - B_y k_x$.
In our model such terms are only allowed if the system is placed on a substrate, while the quadratic terms of Eq.~\eqref{Bsquared} exists regardless of substrate and give the non-linear planar Hall effect.

For the off-diagonal part notice that 
\begin{align}\label{sigma+}
    \sigma_+ = \begin{bmatrix}
        0 & \langle \Phi_{\bm{K}}^c|\mathcal{H}|\Phi^v_{\bm{K}} \rangle \\ 0 & 0 
    \end{bmatrix} 
\end{align}
acquires the phase under the $C_3$ rotation,
$C_3 \sigma_+ = \langle C_3\Phi_{\bm{K}}^c|\mathcal{H}| C_3 \Phi^v_{\bm{K}} \rangle = e^{ i 4 \pi /3} \sigma_+$ and similarly, $C_3 \sigma_- = e^{ -i 4 \pi /3} \sigma_-$.
The immediate consequence of these relations is that the Hamiltonian is diagonal strictly at $\bm{K}$.
To fix the Hamiltonian to linear order in $\bm{k}$ note that the combinations, $k_{\pm} = k_x \pm i k_y$ transform as
$C_3 k_{\pm} = e^{\mp i 2 \pi /3 } k_{\pm}$.
In result, the most general invariant to this order allowed by the group of $\bm{K}$ takes the form, 
\begin{align}\label{HK+}
    H_{\bm{K}}^{(1)}(\bm{k}) = v \sigma_+ k_- + v^* \sigma_- k_+ \, ,
\end{align}
where $v$ is a complex constant. 
This, however is not a finite form of the Hamiltonian as it is further constrained by the $m_x$ and $\mathcal{T}$ symmetries, and this we explore next. 

\begin{table*}[t]
\centering
\caption{The transformation properties of the Pauli matrices parametrizing the Hamiltonian and the momentum counted relative to $\pm \bm{K}$ under the $m_x$ mirror operation and the time-reversal operation, $\mathcal{T}$.
As both operations exchange the valley if the original Pauli matrices parametrize the $H_{\bm{K}}$ the transformed ones parametrize $H_{-\bm{K}}$. Thanks to our choice of the basis, Eq.~\eqref{choice} the Pauli matrices transform trivially in the above sense.    }
\label{tab:m_xT}
\begin{tabular}{|c||c|c|c||c|c||c|c|}
\hline
 &   $\sigma_x$ & $\sigma_y$ & $\sigma_z$ & $k_x$ & $k_y$ & $B_x$ & $B_y$ \\
\hline
         $m_x$  &   $\sigma_x$ & $\sigma_y$ & $\sigma_z$ & $-k_x$ & $k_y$ & $B_x$ & $-B_y$ \\ \hline \hline 
     $\mathcal{T}$  &   $\sigma_x$ & $-\sigma_y$ & $\sigma_z$ & $-k_x$ & $-k_y$ & $-B_x$ & $-B_y$ \\ \hline
\end{tabular}\, .
\end{table*}

The $m_x$ symmetry tells us that all the Pauli matrices such as Eq.~\eqref{sigma+} are the same for the two valies, $\bm{k}=\bm{K}$ and $\bm{k}=-\bm{K}$.
Away from these high symmetry points, based on Table~\ref{tab:m_xT} and \eqref{HK+}, $m_x$  implies 
\begin{align}\label{HK-1}
    H_{-\bm{K}}^{(1)} = -v \sigma_+ k_+ - v^* \sigma_- k_- \, ,
\end{align}
Similarly the time reversal implies,
\begin{align}\label{HK-2}
     H_{-\bm{K}}^{(1)}= -v \sigma_- k_- - v^* \sigma_+ k_+ \, .
\end{align}
The consistency between \eqref{HK-1} and \eqref{HK-2} requires $v$ is a real number.
In summary, we have 
\begin{align}\label{HK-3}
    H_{s\bm{K}}^{(1)}= v ( s \sigma_x k_x + \sigma_y k_y)\, .
\end{align}

Finally we derive the coupling to the in-plane magnetic field, $\bm{B}$.
Such term is requires breaking of the basal mirror symmetry, $m_z$ which is achieved by placing the whole system on a substrate.
As we do not consider the actual spin the coupling to the magnetic field we have in pir problem is via the orbital degrees of freedom in the form of the $\bm{B}\cdot \bm{L}$ coupling.
To have $\bm{B}$ in the effective Hamiltonian a finite admixture of the $m_z$-odd orbitals to the $m_z$-even orbitals $\{ d_{z^2}, d_{x^2-y^2},d_{xy} \}$.
This can only happen if $m_z$ symmetry is broken e.g. by a substrate.
The combination $B_x \pm i B_y$ transforms as $k_x \pm i k_y$.
Therefore, the allowed invariants are real and imaginary parts of $B_+ \sigma_-$.
The preliminary allowed Hamiltonian reads $H^B_{\bm{K}} = \delta (B_x \sigma_x + B_y \sigma_y) + \gamma (B_x \sigma_y - B_y \sigma_x)$.
The $m_x$ symmetry gives 
\begin{align}
    H_{-\bm{K}}^B = \delta  (B_x \sigma_x - B_y \sigma_y) + \gamma (B_x \sigma_y + B_y \sigma_x).
\end{align}
The $\mathcal{T}$ yields 
\begin{align}
    H_{-\bm{K}}^B = \delta  (-B_x \sigma_x + B_y \sigma_y) + \gamma (B_x \sigma_y + B_y \sigma_x).
\end{align}
The consistency of these equations enforces $\delta=0$, and we have 
\begin{align}\label{HB}
    H^B_{s\bm{K}} = \gamma (B_x \sigma_y - s  B_y \sigma_x)\, .
\end{align}

We now summarize the model Hamiltonian that contains Eqs.~\eqref{strain21}, \eqref{HK-3}, \eqref{HB}, and \eqref{diag_H},
\begin{align}\label{H_finite}
    H = s \alpha k_x + v ( s \sigma_x k_x + \sigma_y k_y)+ \gamma (B_x \sigma_y - s  B_y \sigma_x)+\left(m + \frac{\beta k^2}{2} \right)\sigma_z\, .
\end{align}
We note that upon the $\pi/2$ rotation, $k_x \rightarrow -k_y$, $k_y \rightarrow k_x$, and redefining Eq.~\eqref{H_finite} becomes at $B=0$,
\begin{align}\label{H_finite1}
    H = - s \alpha k_y + v (- s \sigma_x k_y + \sigma_y k_x)+ \left(m + \frac{\beta k^2}{2} \right)\sigma_z\, .
\end{align}
This Hamiltonian becomes identical to the one formulated in \cite{Sodemann2015} upon the redefinition $\alpha \rightarrow-\alpha$, at $v_x = v_y = v$.
This explains clearly why in our setting the mirror symmetry preserved by strain is $m_x$ while in the setting describing \eqref{H_finite1} the preserved mirror is $m_y$.
For this reason our BCD is along $\hat{x}$ and in the \eqref{H_finite1} is along $\hat{y}$. 
Correspondingly, in our setting the Hall current from BCD is along $\hat{y}$ for $\bm{E} \parallel \hat{x}$ and in the \eqref{H_finite1} setting along $\hat{x}$ for $\bm{E} \parallel \hat{y}$. 


\end{widetext}

\end{appendix}

\bibliography{shbiblio}

\end{document}